\documentclass{aastex6}

\usepackage{graphicx,epsfig,fancyhdr,psfig,rotating,amsmath,epsf,txfonts,natbib,epstopdf,multirow,natbib}
\usepackage{subfigure}
\usepackage{tabularx}
\usepackage{appendix}
\usepackage{dcolumn}
\usepackage{threeparttable}
\usepackage{longtable}

\def \kms {{\rm km\;s$^{-1}$}}

\def \arcsec {$^{''}$}

\def \mgvi {Mg\,{\sc vi}}
\def \mgvii {Mg\,{\sc vii}}
\def \mgvi {Mg\,{\sc vi}}
\def \sivii {Si\,{\sc vii}}
\def \six {Si\,{\sc x}}
\def \feviii {Fe\,{\sc viii}}
\def \feix {Fe\,{\sc ix}}
\def \fex {Fe\,{\sc x}}
\def \fexi {Fe\,{\sc xi}}
\def \fexii {Fe\,{\sc xii}}
\def \fexiii {Fe\,{\sc xiii}}
\def \fexiv {Fe\,{\sc xiv}}
\graphicspath{{figs/}}
\begin{document}
\title{Plasma parameters and geometry of cool and warm active region loops}
\author{Haixia Xie\altaffilmark{1},
Maria S. Madjarska\altaffilmark{2,1},
Bo Li\altaffilmark{1},
Zhenghua Huang\altaffilmark{1},
Lidong Xia\altaffilmark{1},
Thomas Wiegelmann\altaffilmark{2},
Hui Fu\altaffilmark{1},
Chaozhou Mou\altaffilmark{1}
}

\altaffiltext{1}{Shandong Provincial Key Laboratory of Optical Astronomy and Solar-Terrestrial Environment, Institute of Space Sciences, Shandong University, Weihai, 264209 Shandong, China}
\altaffiltext{2}{Max Planck Institute for Solar System Research, Justus-von-Liebig-Weg 3, 37077, G\"ottingen, Germany}


\date{Received date, accepted date}

\begin{abstract}
How the solar corona is heated to high temperatures remains an unsolved mystery in solar physics. In the present study we analyse
observations of 50 whole active-region loops taken with the Extreme-ultraviolet Imaging Spectrometer (EIS) on board the
Hinode satellite. Eleven loops were classified as cool ($<$1~MK) and 39 as warm (1--2~MK) loops. We study their plasma parameters such as densities, temperatures, filling factors, non-thermal velocities and Doppler velocities. We combine spectroscopic analysis with linear force-free magnetic-field extrapolation to derive the three-dimensional structure and positioning of the loops,  their lengths and heights  as well as the magnetic field strength along the loops. We use density-sensitive line pairs from Fe~{\sc xii}, Fe~{\sc xiii}, Si~{\sc x} and Mg~{\sc vii} ions to obtain electron densities by taking  special care of intensity background subtraction. The emission-measure loci method is used to obtain the loop temperatures. We find that the loops are nearly isothermal along the line-of-sight. Their filling factors are between 8\% and 89\%. We also compare the observed parameters with the theoretical RTV scaling law. We find that most of the loops are in an overpressure state relative to the RTV predictions. In a followup study, we will report a heating model of a parallel-cascade-based mechanism and  will compare the model parameters  with the loop plasma and structural parameters derived  here.

\end{abstract}


\maketitle

\section{Introduction}
\label{sect_intro}
Loops are one of the main building blocks of the solar atmosphere. Understanding their heating, however, remains a huge challenge \citep{2006SoPh..234...41K}. Particularly, the question on whether the plasma is heated by steady or impulsive, uniform or localized mechanism(s) is still open \citep{2010ApJ...709..499S}. \citet{1998Natur.393..545P} noted that the physical parameter (e.g., temperature and density) profiles along a coronal loop are highly sensitive to the heating mechanisms. Accordingly, theoretical heating models require accurate measurements of the plasma parameters along coronal loops, e.g. temperature, density, filling factor, velocity, magnetic field, etc.

\par
A number of authors have used imaging or spectral data  to investigate the physical properties of coronal loops. Based on observations, loops are divided into cool, warm, and hot loops. Cool loops are those typically detected in Ultraviolet (UV)/Extreme Ultraviolet (EUV) lines with formation temperatures between 0.1 and 1~MK (e.g., \citealt{2014LRSP...11....4R, 2015ApJ...800..140G}). The temperature range of warm loops is 1--2~MK (e.g., \citealt{1999ApJ...517L.155L}). Hot loops are usually observed in soft X-rays. They are located in active regions  and have temperatures higher than 2~MK (e.g., \citealt{1995PASJ...47L..15Y}).
\begin{table*}[!ht]
\centering
\caption{EIS data used in the present study.}\label{EIS DATA}
\begin{tabular}{c c c c c c c c}
\hline
\hline
DataNo.(DaNo.)&Date & NOAA &Observational&(X-cen, Y-cen) &Exp& Slit&Lines and channel \\
LoopNo.(LpNo.)&&&time (UT)&(arcsec)&time (s)&(arcsec)&(EIS, AIA) \\
\hline
Da1 Lp1&2007-01-16 &10938& 01:54 -- 02:20 & (-490, 35) & 5 & 1& \fexi, \fexii, \fexiii, \fexiv,193~\AA\  \\
Da2 Lp2&2007-09-30 &10971& 06:50 --  08:01 & (168, -75) & 15 & 1&\feviii, \fex, \sivii, 171~\AA\ \\
Da3 Lp3, 4, 5&2007-10-07 &10972& 08:21 -- 08:52 & (195, -202) & 30 & 2&\feviii, \fex, \mgvii, \sivii, 171~\AA\ \\
Da4 Lp6&2007-10-08 &10972& 02:23 -- 02:54 & (338, -207) & 30 &2& \feviii, \fex, \mgvii, \sivii, 171~\AA\  \\
Da5 Lp7&2010-06-18 &11082& 19:21 --  19:47 & (-315, 448) & 30 & 2&\feviii, \mgvi, \mgvii, \sivii, 171~\AA\ \\
Da6 Lp8&2010-09-14 &11106& 12:00 -- 13:18 & (-593, -422) & 30 & 2&\six, \fexii, \fexiii,\fexiv, 193~\AA\  \\
Da7 Lp9&2010-09-22 &11108& 23:05 -- 00:24 & (34, -567) & 30 & 2&\six, \fexii, \fexiii, \fexiv, 193~\AA\ \\
Da8 Lp10&2010-09-29 &11109& 22:32 -- 01:10 & (374, 225) & 30 & 1&\feviii, \fex, \mgvii, \sivii, 171~\AA\  \\
Da9 Lp11&2010-11-30 &11130& 09:53 --  11:23 & (299, 183) & 15 & 1&\fex, \fexii, \fexiv, 193~\AA\ \\
Da10 Lp12&2011-02-06 &11150& 22:40 -- 23:58 & (734, -310) & 40 & 2&\feviii, \fex, \mgvii, 171~\AA\  \\
Da11 Lp13&2011-03-19 &11175& 08:28 -- 10:16 & (303, 257) & 60 & 1&\feviii, \feix, \fex, \mgvii, \sivii, 171~\AA\ \\
Da12 Lp14, 15&2011-04-15 &11190& 00:15 -- 02:18 & (212, 351) & 60 & 1&\fex, \fexi, \fexii, \fexiii, \fexiv, 193~\AA\ \\
Da13 Lp16, 17&2011-04-21 &11193& 11:06 -- 13:09 & (459, 323) & 50 &1& \fex, \fexi, \fexii, \fexiii, 193~\AA\  \\
Da14 Lp18&2011-06-14 &11234& 11:34 -- 12:52 & (55, -255) & 30 &2&\fexii, \fexiii, \fexiv, \six, 193~\AA\  \\
Da15 Lp19&2011-06-15 &11234& 00:40 -- 01:58 & (167, -253) & 30 & 2&\fexii, \fexiii, \fexiv,\six, 193~\AA\  \\
Da16 Lp20&2011-12-06 &11363& 09:56 -- 11:59 & (198, -387) & 60 & 1&\feix, \fex, \fexi, \fexii, \fexiii, 193~\AA\  \\
Da17 Lp21&2012-01-13 &11395& 16:30 -- 18:33 & (-19, 323) & 60 & 1&\fexi, \fexii, \fexiii, \fexiv, 193~\AA\ \\
Da18 Lp22&2012-03-28 &11445& 10:20 -- 13:25 & (-284, -309) & 60 & 1&\feix, \fex, \fexi, \fexii, 193~\AA\  \\
Da19 Lp23&2012-08-23 &11548& 07:37 -- 08:23 & (-283, 220) & 45 &2& \fex, \fexi, \fexii, \fexiii, 193~\AA\  \\
Da20 Lp24&2012-09-25 &11575& 20:21 -- 21:31 & (190, 50) & 40 & 1&\fex, \fexi, \fexii, \fexiii, 193~\AA\  \\
Da21 Lp25&2012-12-19 &11633& 18:13 -- 20:16 & (-280, -100) & 60 & 1&\fexi, \fexii, \fexiii, \fexiv, 193~\AA\ \\
Da22 Lp26&2012-12-23 &11633& 09:30 --  11:33 & (522, -107) & 60 & 1&\feviii, \fex, \mgvii, \sivii, 171~\AA\ \\
Da23 Lp27&2013-01-16 &11654& 12:02 -- 13:07 & (488, 183) & 30 & 2&\fex, \fexi, \fexii, \fexiii, \fexiv, 193~\AA\ \\
Da24 Lp28&2013-06-10 &11765& 11:00 -- 12:01 & (622, 133) & 35 &2&\fex,  \fexi, \fexii, \fexiii, 193~\AA\  \\
Da25 Lp29&2013-08-08 &11809& 16:08 -- 17:13 & (480, 129) & 30 &2&\fex, \fexi, \fexii, \fexiii, \fexiv, 193~\AA\  \\
Da26 Lp30&2014-01-26 &11960& 10:03 -- 12:06 & (361, -309) & 60 & 1&\fex, \fexi, \fexii,\fexiii, 193~\AA\  \\
Da27 Lp31&2014-02-18 &11976& 23:50 -- 01:53 & (456, -54) & 60 & 1&\fex, \fexi, \fexii, \fexiii, 193~\AA\  \\
Da28 Lp32&2015-01-08 &12257& 13:35 -- 15:38 & (93, 250) & 60 & 1&\fex, \fexi, \fexii, \fexiii, \fexiv, 193~\AA\  \\
Da29 Lp33&2015-03-23 &12303& 18:13 --  19:14 & (107, 303) & 60 & 1&\fexi, \fexii, \fexiii, \fexiv, 193~\AA\ \\
Da30 Lp34&2015-05-22 &12351& 10:28 -- 11:29 & (-280, 386) & 60 & 1&\fex, \fexi, \fexii, \fexiii, \fexiv, 193~\AA\ \\
Da31 Lp35,36&2015-06-10 &12362& 00:26 -- 01:35 & (-34, 87) & 40 & 2&\fex, \fexi, \fexii, \fexiii, 193~\AA\  \\
Da32 Lp37&2015-06-11 &12362& 01:04 -- 02:13 & (195, 81) & 40 & 2&\fex, \fexi, \fexii, \fexiii, 193~\AA\  \\
Da33 Lp38&2015-07-09 &12381& 10:40 -- 12:43 & (33, 124) & 60 & 1&\fex, \fexi, \fexii, \fexiii, \fexiv, 193~\AA\  \\
Da34 Lp39&2015-07-28 &12390& 21:12 -- 22:13 & (363, -396) & 60 & 1&\fex, \fexi, \fexii, \fexiii, 193~\AA\  \\
Da35 Lp40&2015-08-09 &12369& 01:50 --  02:51 & (174, -440) & 60 & 1&\fex, \fexi, \fexii, \fexiii, 193~\AA\ \\
Da36 Lp41, 42&2015-09-03 &12409& 08:23 -- 09:33 & (-170, -437) & 40 & 2&\fex, \fexi, \fexii, \fexiii, \fexiv, 193~\AA\ \\
Da37 Lp43&2015-09-04 &12410& 08:59 -- 10:09 & (74, -437) & 40 &2&\fex, \fexi, \fexii, \fexiii, \fexiv, 193~\AA\  \\
Da38 Lp44&2015-12-25 &12473& 03:32 -- 04:41 & (-513,-338) & 40 & 2&\fex, \fexi, \fexii, \fexiii, 193~\AA\  \\
Da39 Lp45&2016-01-12 &12483& 22:50 -- 23:51 & (116, 483) & 60 & 1&\fex, \fexi, \fexii, \fexiii, 193~\AA\  \\
Da40 Lp46, 47&2016-06-28 &*& 02:44 -- 04:49 & (-551, 88) & 45 & 2&\fex, \fexi, \fexii, \fexiii, 193~\AA\  \\
Da41 Lp48&2016-07-17 &12564& 20:44 -- 21:37 & (619, 125) & 30 & 2&\feix, \feviii, \fex, \sivii, 171~\AA\ \\
Da42 Lp49, 50&2016-07-19 &12567& 11:52 -- 12:57 & (364, -17) & 30 & 2&\fex, \fexi, \fexii, \fexiii, 193~\AA\  \\
\hline
\hline

\end{tabular}
\end{table*}

\begin{table}[!ht]
\centering
\caption{EIS spectral lines.}\label{spectral line}
\begin{tabular}{c c c}
\hline
\hline
Ion &$ \lambda~({\AA})$& logT$_{max}$~(K) \\
\hline
\mgvi&268.99&5.7 \\
\mgvii&278.40&5.8\\
       &280.74&5.8\\
\sivii&275.36&5.8\\
\six&258.37&6.1\\
    &261.04&6.1\\
\feviii&186.60&5.8\\
\feix&188.49&5.9\\
\fex&184.54&6.0\\
     &257.26&6.0\\
\fexi&188.23&6.0\\
\fexii&186.88&6.1\\
       &192.39&6.1\\
       &195.12&6.1\\
\fexiii&202.04&6.2\\
&203.83&6.2\\
\fexiv&264.79&6.3\\

\hline
\end{tabular}
\end{table}

\par
Warm loops are reported in several comprehensive studies based on spectral data taken with the Coronal Diagnostic Spectrometer (CDS) on board the Solar and Heliospheric Observatory (SoHO) \citep[e.g.,][]{2001ApJ...556..896S, 2003A&A...406.1089D, 2008SoPh..252..293S} and Extreme-ultraviolet Imaging Spectrometer (EIS) on board the Hinode satellite \citep[e.g.,][]{2009ApJ...694.1256T, 2008ApJ...686L.131W, 2012SoPh..276..113S, 2015ApJ...800..140G}. There are also a number of studies on warm loops using imaging observations  \citep[e.g.,][]{1999ApJ...517L.155L, 2008ApJ...680.1477A}.

\par
\citet{1999ApJ...517L.155L} reported the temperature and emission measure along segments of four warm loops using imaging data from the Transition Region and Coronal Explorer (TRACE) satellite and compared the observed loop structure with theoretical isothermal and non-isothermal static loop models. They found that the loop temperature profile is near-constant and incompatible with the theoretical results. \citet{2001ApJ...556..896S} studied an active region (AR) warm loop in CDS data and concluded that the temperature distribution is ``inconsistent with isothermal plasma along either the line-of-sight or the length of the loop''. The authors suggested that both radiative and conductive losses are important in the case of their loop. \citet{2009ApJ...694.1256T} obtained the physical parameters along a segment from an AR loop using EIS  data and obtained temperatures of 0.8 -- 1.5 MK, electron densities in the range $10^{9}-10^{8.5}$~cm$^{-3}$ and filling factor from  0.02 to 4 as the loop height increased. They also concluded that the loop is close to isothermal for each position along the loop after accounting for the background emission (see also similar results by \citet{2003A&A...406L...5D} and \citet{2003A&A...406.1089D}).

\par

There are also reports on the physical parameters along the whole length of a loop. For example, \citet{1998Natur.393..545P} analysed a whole loop using Yohkoh Soft X-ray Telescope (SXT) data and concluded that the uniform heating model can fit well the loop observational temperature distribution. More importantly, it was found that the ``observed variation in temperature along a loop is highly sensitive to the spatial distribution of the heating". \citet{2004ApJ...608.1133L}  made use of spectral data from CDS and compared them with a one-dimensional, time-independent, non-static model. They found no agreement between the model predictions and the observations for the whole loop. \citet{2008ApJ...680.1477A} used a triangulation method to obtain the 3D reconstructions of 30 coronal loops in an active region observed simultaneously with the Extreme-Ultraviolet Imaging (EUVI) telescopes on the STEREO A and B spacecrafts and adopted an emission measure method from triple-filter images to derive the densities and temperatures of the loops. They compared the obtained pressure of all loops with   the pressure predicted by the Rosner-Tucker-Vaiana (RTV) scaling law \citep{1978ApJ...220..643R} and found that the loops were in a ``typical overpressure state''. \citet{2015ApJ...800..140G} reported the analysis of an entire loop observed with Hinode/EIS and found that the coronal loop is not in hydrostatic equilibrium state.

\par
 In the present study, apart from carrying out a comprehensive spectral diagnostics of 50 whole loops observed in  42 active regions, we combine for the first time such diagnostics with obtaining the 3D magnetic-field structure and magnetic field strength along each loop  using a linear force-free field extrapolation. The large whole loop selection permits  to statistically compare with the RTV scaling law. Moreover, the combination of magnetic-field information and the physical parameter information  of the  coronal loops  will then be used to test a loop heating model (to be reported in a follow-up paper). We describe the observational data in Section~\ref{sect_obs}. The methodology is introduced in Section~\ref{sect_met}. The results and discussion are given  in Section~\ref{results}. In Section~\ref{sum} we summarize our results.

\section{DATA}
\label{sect_obs}
 In order to study the physical parameters of coronal loops, we searched the EIS \citep{2007SoPh..243...19C} archive  for suitable spectroscopic data.  Fifty AR loops  were found in forty two active regions. Line-of-sight magnetograms taken with the Helioseismic Magnetic Imager \citep[HMI,][]{2012SoPh..275..229S} and the  Michelson Doppler Imager \citep[MDI,][]{1995SoPh..162..129S} were used for a linear force-free  magnetic-field extrapolation. Details on the observations are given below. Data from the Atmospheric Imaging Assembly \citep[AIA,][]{2012SoPh..275...17L} on board the Solar Dynamic Observatory (SDO) and the  Extreme-ultraviolet Imaging Telescope \citep[EIT,][]{1995SoPh..162..291D} on board SoHO are then used to cross-check the extrapolation.

\subsection{Spectroscopic Data}
Table~\ref{EIS DATA} gives information about the EIS data including the observational dates, the NOAA (National Oceanic and Atmospheric Administration) active region numbers, the field-of-views (FOVs), X(Y)-center in arcseconds, the exposure times, the slit sizes, the EIS spectral lines and the AIA channels used in the present analysis. The EIS pixel size is 1\arcsec\ and the spectral sampling is 22~m\AA$/$pixel. The analysed spectral lines  are listed in Table~\ref{spectral line}. The EIS data were processed with the standard software.  The data were calibrated to intensity units $ \mathrm{erg~cm^{-2} s^{-1} sr^{-1} ~\AA\ ^{-1}}$. There are two instrumental effects that need to be corrected: the tilt of the grating dispersion axis relative to the CCD axes and the spatial offset in Y direction between the two CCDs, which leads to offsets along the slit direction and the raster-scanning direction between the images taken by the two CCDs. In our study, an offset in the range  17\arcsec\ -- 18\arcsec\ along the slit (solar-Y direction) was estimated  between the short wavelength and the long wavelength channels. The  He~{\sc ii}~256.32~\AA\ and \feviii\ 186.60~\AA\ lines were used for the Y offset calculations in the case of the cool loops, and the He~{\sc ii}~256.32~\AA\ and \fexii\ 195.12~\AA\ lines for the warm loops.  An offset of 1\arcsec\ in the solar-X direction was obtained. The procedure \textit{eis\_wave\_corr.pro} was used to correct the spectrum drift that is caused  by thermal effects related to the satellite's orbit around the Earth \citep{2010SoPh..266..209K}. The \textit{eis\_auto\_fit.pro} procedure was used to fit the spectral lines with a Gaussian model.

\begin{figure}[!ht]
\centering
\includegraphics[scale=0.55]{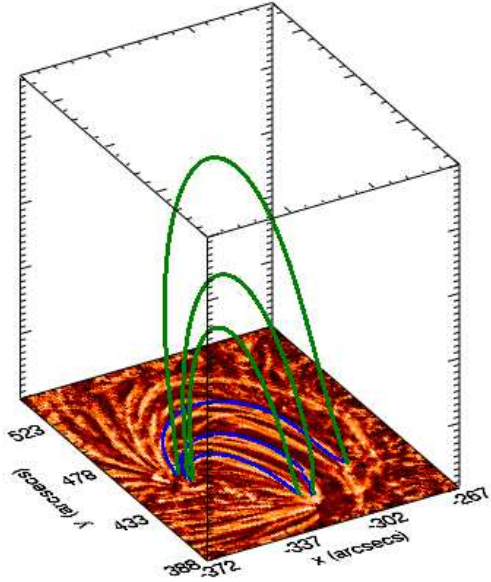}
\caption{Extrapolated results shown in a 3D coordinate system.  The Z-axis has been stretched. The X-Y plane image represents a MGN enhanced AIA image with overplotted blue lines that are the projections of the extrapolated loops shown with green lines.}
\label{3d_plot}
\end{figure}

\subsection{Magnetograms and Imaging Data}

HMI provides line-of-light (LOS)  magnetograms at a spatial sampling of 0.5\arcsec\ per pixel and a cadence of 45~s. In the present work, we performed a linear force-free field extrapolation to obtain the three-dimensional (3D) topology of the loops \citep{2005SoPh..228...67W}. The magnetic field is calculated in a rectangular coordinate system for the regions located near the center of the Sun. However, for  regions located away from the disk center it is necessary  to implement a curvature correction \citep{1990SoPh..126...21G}. First, the LOS magnetograms data were corrected for the projection effect, i.e. from B$_{LOS}$ (line-of-sight magnetic flux) to B$_{r}$ (radial magnetic flux). Next,  B$_{r} $ in the solar-disk coordinate system is converted to a spherical coordinate system by bilinear interpolation which is then used as the initial magnetic field value for the extrapolations. The extrapolated results were then converted for convenience to the image coordinate system  to compare with the imaging and spectral data. Details about the transformation between the image coordinate system and the heliographic coordinate system can be found in \citet{1999ApJ...515..842A}.

\par
The AIA imager takes full-disk images in 12 EUV and UV channels with a spatial sampling of  0.6\arcsec\ per pixel and a temporal cadence of 12~s (the UV and visible channels are taken at a lower cadence). We used only the 171~\AA\ (logT = 5.8~K) and 193~\AA\ (logT = 6.2~K) channels. The alignment of AIA and EIS was done by comparing solar features using EIS intensity images taken in spectral lines with formation temperatures close to the AIA channels. We processed the AIA data by enhancing the loop structures using the Multi-Scale Gaussian Normalisation (MGN) method developed by \citet{2014SoPh..289.2945M} in order to better compare with the magnetic-field lines derived from the magnetic-field extrapolation.

We used observations from MDI and EIT for the datasets taken before 2010, i.e. Da39--42. The MDI observations were taken with a FOV of 1024 $\times$ 1024 pixels$^2$  and a $\approx$2\arcsec\ per-pixel spatial resolution. The EIT data have a pixel size of 2.6\arcsec and were taken with the Fe~{\sc XII} 195~\AA\ filter.

\section{Methodology}
\label{sect_met}

In this section we will describe the methodology used in the present paper on the linear force-free extrapolation, the electron density and temperature diagnostics, the filling factor estimation as well as the LOS velocities, the velocities along the loops (only for one loop, see further in the text), and the non-thermal velocities.

\label{geo}
Presently, direct and accurate measurements of the magnetic field in the corona are very challenging. However, magnetic-field extrapolations from magnetic field measurements taken at photospheric or chromospheric levels are feasible if we make assumptions about current densities in the corona. The condition of a low average plasma-$\beta$ allows us to assume that to the lowest order the magnetic field is force-free, i.e. the current density is aligned with the magnetic field. More detailed information about the method can be found in \citet{2005SoPh..228...67W}. We normalize the linear force-free parameter $\alpha$ by the harmonic mean
L  of L$_{x}$ (the rectangular x length, if the magnetogram  represents a rectangular region)
and L$_y$ (the rectangular y length) of the magnetogram defined by 1/L$^2$=1/2(1/L${_x}^2$ + 1/L${_y}^2$).
In the paper we use the dimensionless quantity $\alpha$L.   We selected one value of $\alpha$L that agrees best with at least two loops observed in the AIA and EIS observations. The loop best visible in the EIS images was used for a further spectroscopic analysis. From the extrapolation we obtained the 3D loop structure, its length, height and the magnetic field distribution along the loop length. Then the angle $\theta$ between the LOS and the loop arc direction was stipulated in order to obtain the plasma velocity along the loop. In our analysis, the positive direction of the loop is from the east to the west footpoint of the loop. A 3D example of the extrapolated results for one of the datasets is shown in Figure~\ref{3d_plot}. It presents three observational loops that conform well with the extrapolated magnetic field. As mentioned above, only the loop best recorded in the EIS data was used for further analysis.

\par

To obtain the electron density, we used the intensity ratios of spectral lines of the same ion which are sensitive to electron density. The theoretical intensity  ratio was calculated using CHIANTI version 7.0 \citep{1997A&AS..125..149D, 2012ApJ...744...99L, 2013ApJ...763...86L}. Several EIS spectral line pairs were selected for density diagnostics as suggested by \citet{2007PASJ...59S.857Y, 2009A&A...495..587Y}. In our analysis, we used four spectral line pairs depending on the temperature of each loop, i.e., the visibility of each loop in these spectral lines.  After a thorough  analysis regarding the signal-to-noise ratio,  we selected only line profiles with peak intensities above 120 erg~cm$^{-2}$~s$^{-1}$~sr$^{-1}$~\AA$^{-1}$. Mostly affected were the Mg~{\sc vii} lines, and the pixels with a poor signal were removed from the analysis.  The \mgvii\ $278.40/280.74$~\AA\ pair was analysed for cool loops, while the ratios of \fexii\ $186.88/(195.12+195.18)$~\AA, \fexiii\ $202.04/(203.80+203.83)$~\AA\, and \six\ $258.37/261.04$~\AA\ were used for warm loops. In order to increase the signal-to-noise ratio, we summed the signal from $3pixels \times 3pixels$ boxed regions along the loop.  The Gaussian fitting error is considered as an important aspect for the density error estimation and the spectral line blending issues were  also taken into account. For the correctness of the diagnostics, the intensity background subtraction is very important as it affects directly the outcome of the results \citep{2002A&A...383..952R, 2000ApJ...539.1002P, 2001ARA&A..39..175A}.  In the present study, the background emission was obtained as follows.
To start, we constructed an artificial transverse slit at each selected location along the loop, thereby obtaining a cross-loop intensity profile.
Examining these profiles, we found that either no obvious structure can be seen in the neighbourhood of the loop or
   only one nearby structure exists.
The location (labeled ``b'' hereafter) where we identify the intensity as the background emission was obtained by distinguishing between these two possibilities.
In the first case, we found the locations where the cross-loop intensity profile starts to level off, typically finding
   two on opposite sides.
Location ``b'' was chosen to be the one with lower intensity (see the majority of the blue boxes in the leftmost panel of Fig.~\ref{t1}).
When the second possibility arises, location ``b'' was chosen to be the point where the intensity profile levels off
   on the side where no nearby structure exists (see the blue boxes close to the east footpoint in Fig.~\ref{t1}).
Finally, a $3pixels \times 3pixels$ box was constructed with ``b'' at the center, and the signals were averaged over this boxed region
   to increase the signal-to-noise ratio of the background emission.
We note that at some pixels along some loops, the densities we derive may differ from the one derived at adjacent pixels to such an extent
   that we think the background emissions cannot be reliably determined.
These pixels were excluded from any further diagnostics
   (see the portion close to the west footpoint in Figs.~\ref{tn2}a).

\par
We derived the electron temperatures using the Emission-Measure (EM)-loci method (\citet{1987MNRAS.225..903J} and \citet{2002A&A...385..968D}). For the contribution function G(T) calculations we used the elemental abundances from \citet{1992PhyS...46..202F}.
If we assume that the temperature is isothermal along the LOS,  the EM curves will cross at a single location. The temperature is then obtained by averaging  the values for the close-by points of intersection of the EM  curves and the standard deviation is taken as the error of the temperature estimation.

 In our analysis, the method used by \citet{2008ApJ...686L.131W} was applied to calculate the filling factor along the coronal loops. Under the assumption that the loop emitting volume is a cylinder, the LOS emission measure can be simplified by dividing the volume emission measure by the area of an EIS pixel and is thus written as:
\begin{equation}
\mathrm{EM=f\frac{n_{e}^{2}\pi r^{2}}{l}},
\label{equat5}
\end{equation}
where f is the volumetric filling factor, r is the observed radius of the loop, and l is the length of the EIS pixel. To obtain the observed radius we fit the sliced intensity across loop points that are well visible. The full width of the half maximum (FWHM) obtained after a Gaussian fit is taken as the diameter of the loop. For each loop, the filling factor was estimated only in  points along the loops where the radius measurements  were reliable. The radii were determined only for one spectral line where the loops were registered with the best signal-to-noise ratio.

\par
The velocities along the length of the loop  $\mathrm{V_{loop}}$  based on the Doppler velocities ($\mathrm{V_{D}}$)  and the 3D coordinate information from the extrapolation were also calculated (only for one loop, see below). They were obtained as $\mathrm{V_{loop} = -V_{D}/\cos{\theta}}$ ($\mathrm{\theta}$ is the angle between the LOS and the loop direction).  The positive direction of the velocity is  from the east footpoint to  the west footpoint of the loop.  In our analysis, the velocities along the loop are removed when $\mathrm{| \theta - 90^{\circ} | < 15^{\circ}}$ because this range of angles produces large and meaningless values.

\par
Under the assumption that the spectral lines are Gaussian, the measured line widths consist of three components: an instrumental broadening, a thermal broadening, and a non-thermal broadening. The latter may  result from wave and turbulent motions, and is given by

\begin{equation}
\mathrm{(FWHM)_{NT}^{2} = (FWHM)_{Obs}^{2}-(FWHM)_{Ins}^{2}-4\ln2\frac{\lambda^{2}}{c^{2}}\frac{2k_{B}T}{M}},
\label{equat7}
\end{equation}
where $\mathrm{M}$ is the mass of the atom emitting the line, $\mathrm{k_B}$ is the Boltzmann constant, $\mathrm{\lambda}$ is the spectral line wavelength, and  $\mathrm{T}$ is the temperature of the atom that is taken to be equal to the electron temperature of the spectral line given in Table~2, i.e. the maximum of the formation temperature (for more details see \citet{1998ApJ...505..957C}). We obtained the instrumental broadening using the procedure \emph{eis\_slit\_width.pro}. The non-thermal velocity $\mathrm{V_{nt}}$ is obtained as

\begin{equation}
 \mathrm{V_{nt}=(FWHM)_{NT}\cdot c / (\lambda \cdot 2\sqrt{\ln2})}.
\label{equatfwhm}
\end{equation}
It should be noted that  the measurements of the non-thermal velocities in the case of the \fexii~195.12~\AA\ line will be affected by the blend from the \fexii~195.18~\AA\ line. This blend can make the values of the non-thermal velocities smaller, in regions where the density is higher, for example, near the loop footpoints. Details on this blend and its implications can be found in, e.g. \citet{2009A&A...495..587Y} and \citet{2016ApJ...820...63B}.

\par

It has been customary to examine how the density distribution along
   a loop deviates from a hydrostatic equilibrium
   \citep[e.g.,][]{2008ApJ...680.1477A, 2015ApJ...800..140G}.
In its simplest form, this equilibrium comes from the balance
   between gravity and the pressure gradient force.
Given that the electron temperatures ($T$) do not show a significant
   spatial dependence in the vast majority of the loops we examine,
   this pressure gradient force comes primarily from the spatial variation of the electron density ($n$).
As a result, $n$ is simply given by
\begin{equation}
\frac{n}{n_{\mathrm{ref}}}
   = \exp\left(-\frac{\overline{Z}-\overline{Z}_\mathrm{ref}}{60\overline{T}_{6}}\right),
\label{equathy}
\end{equation}
   where $\overline{Z}$ is the loop height in unit of $\mathrm{10^{6}}$~m,
   and $\overline{T}_{6}$ is the temperature (in~MK)
   averaged over a loop.
Note that for simplicity, we have assumed an electron-proton gas.
As for the reference values (subscript $\mathrm{ref}$), we choose the footpoint where
   the density measurement has a lower uncertainty than at the other end.

A few words seem necessary regarding the assumptions behind Equation~(\ref{equathy}),
   which is a simplified version of the momentum equation projected onto a loop.
A more complete version reads
\begin{equation}
\displaystyle
   \hat{l}\cdot\left[\rho \frac{{\mathrm d} {\bf v}}{{\mathrm d} t}\right]
     = - \hat{l}\cdot\nabla P
       - \rho g \hat{l}\cdot\hat{z}
       + \hat{l}\cdot{\bf F}~,
   \label{eq_momentum}
\end{equation}
   where $\rho$ is the mass density, ${\bf v}$ is the velocity,
   $P$ is the thermal pressure,
   and $g= 27400$~cm~s$^{-2}$ is the gravitational acceleration of the Sun.
Furthermore, $\hat{l}$ and $\hat{z}$ are the unit vectors along the loop and
   in the vertical direction, respectively.
In addition, ${\bf F}$ represents some unspecified volumetric force, which may derive
   from, say, Alfv\'en waves~\citep[e.g.,][]{2003ApJ...598L.125L,2012ApJ...746...81A}.
Evidently, Equation~(\ref{eq_momentum}) reduces to Equation~(\ref{equathy})
   when the inertial force
   ($\rho {\mathrm d}{\bf v}/{\mathrm d}t$)
   and ${\bf F}$ are neglected.
As a consequence, the deviation of the measured density profiles from the predictions
   by Eq.~(\ref{equathy}) will be an indication that these additional forces
   can be comparable in magnitude with the pressure gradient force.

\section{Results and Discussion}
\label{results}

In the present study we identified  50 loops observed along their full length in the EIS raster data.  Eleven loops were classified as cool loops (T $<$ 1~MK) and 39 as warm loops (T $\sim$ 1 -- 2~MK). The obtained plasma parameters of the loops as well as their geometry, length, height and the magnetic-field distribution along their full length are reported below and are summarized in Table~\ref{parameters}. The figures of all loops  can be found here \citet[][\url{https://doi.org/10.5281/zenodo.580645}]{haixia_xie_2017_loops}.
\subsection{Cool Loops}

Eleven  cool loops were found in the selected datasets. These are  Loop2, Loop3, Loop4, Loop5, Loop6, Loop7, Loop10, Loop12, Loop13, Loop26 and Loop48.  In the main text we will only give the figures related to Loop7 and Loop12.
\subsubsection{Loop7}
\par
\begin{center}
\begin{figure*}[!ht]
\begin{tabular}{cc}
\includegraphics[width=.60\textwidth]{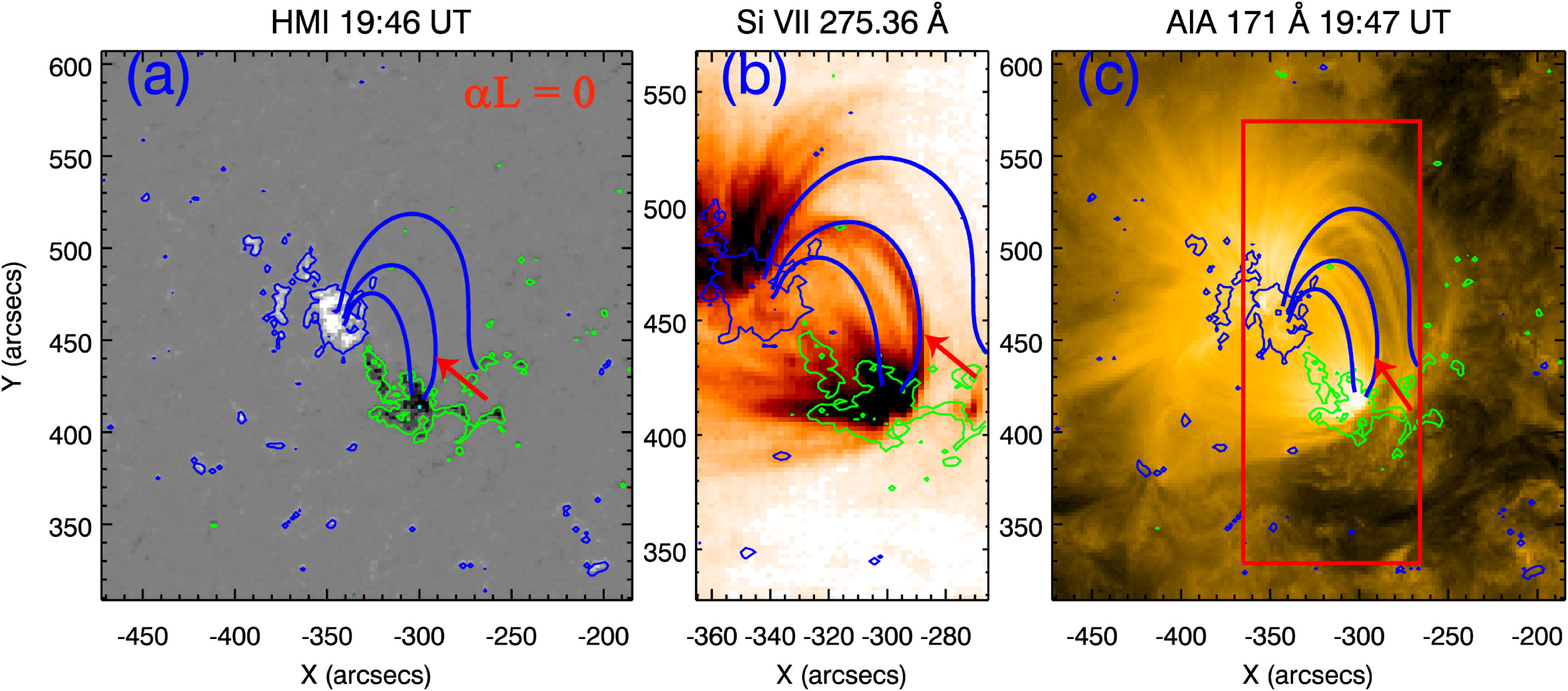} &
\includegraphics[width=.38\textwidth]{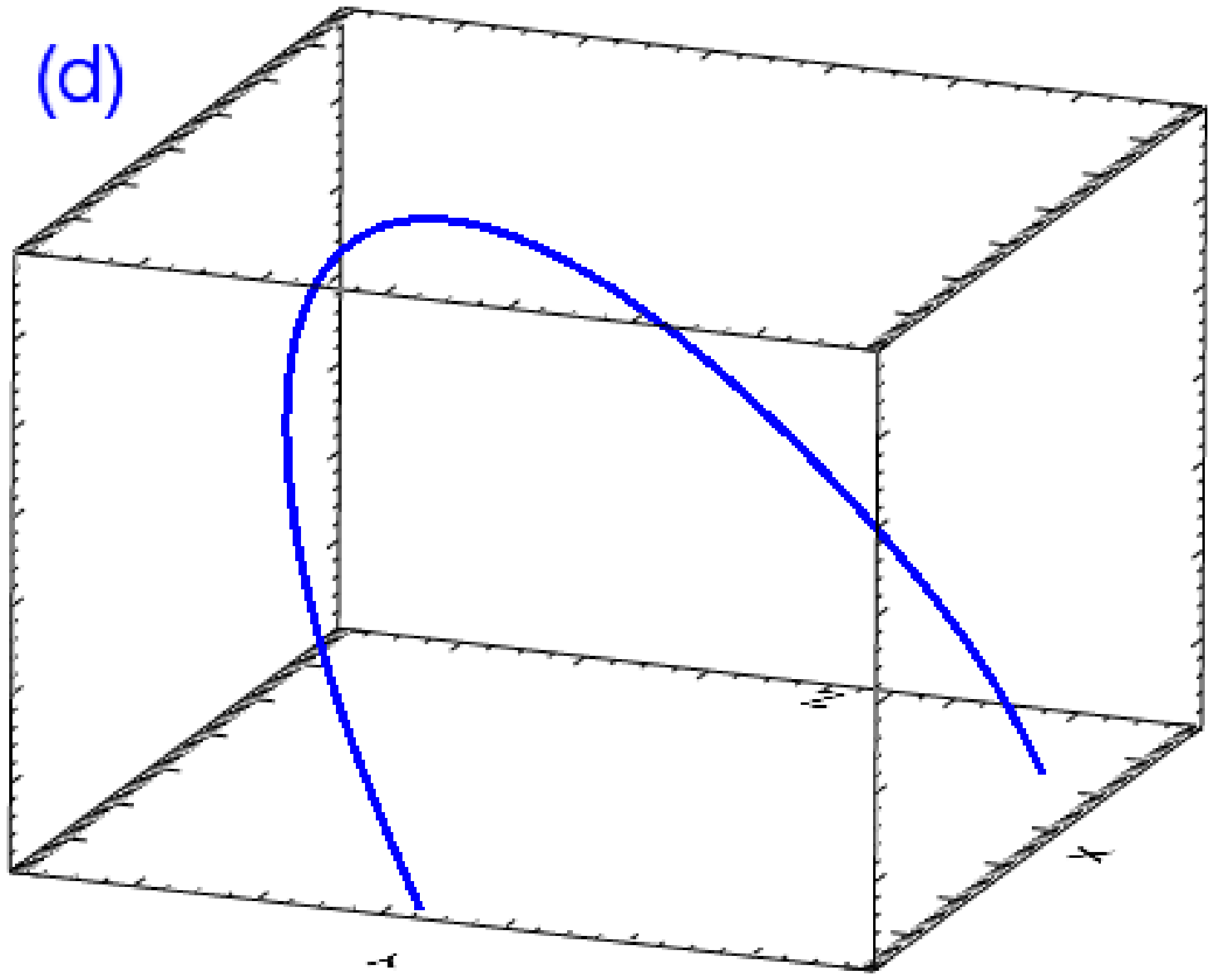}
\end{tabular}
\caption{Cool loop (Loop7) on 2010 June 18. (a): HMI magnetogram scaled from $-$500 to 500~G overplotted with three extrapolated magnetic-field lines and the $\pm$80~G magnetic-field contours. (b): EIS intensity image in the \sivii~275.36~\AA\ line overplotted with the magnetic field-contours and three extrapolated magnetic-field lines. (c): {AIA} 171~\AA\ image overplotted with three extrapolated field lines and  $\pm$80~G magnetic-field contours. The red arrow points at the analyzed loop. The red square is the EIS FOV. (d):  3D view of the extrapolated loop. The axis units are `arcsec'.   }\label{lfff1}
\end{figure*}
\end{center}

Figure~\ref{lfff1}(a) shows the SDO/HMI magnetogram of the loop in NOAA 11082 observed on 2010 June 18 with the extrapolated magnetic-field lines overplotted. We selected three magnetic-field lines with $\alpha L = 0$ that agree  with 3 loops seen in the EIS \sivii\ 275.36~\AA\ raster image (Figure~\ref{lfff1}(b)).  In  Figure~\ref{lfff1}(c) an AIA~171~\AA\ image reveals a more complex view of the loops thanks to the AIA higher spatial resolution  (we come back to this later).  As only one loop is visible along its full length in the EIS \sivii\ 275.36~\AA\ image, this loop was selected for further analysis (noted with a red arrow in Figure~\ref{lfff1}(a), (b) and (c)). The 3D view of the loop is given in Figure~\ref{lfff1}(d) showing the spatial positioning of the loop.
\par
\begin{figure*}[!ht]
\centering
\includegraphics[trim=0cm 0cm 0cm 0cm,clip,width=1.\columnwidth]{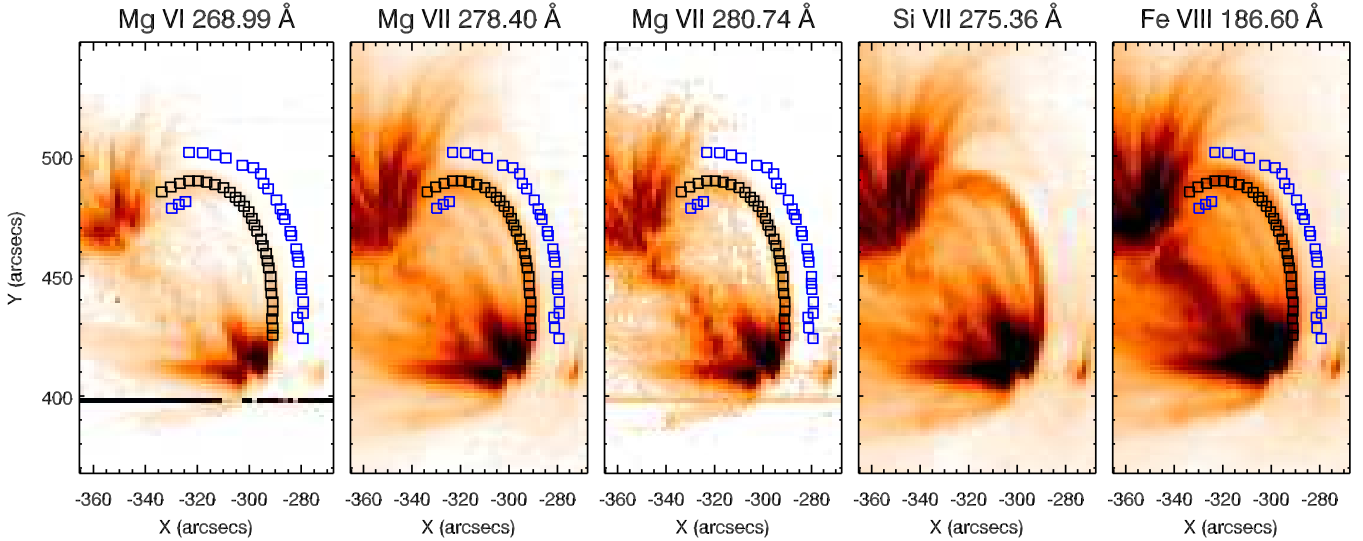}
\caption{Cool Loop7 as seen in different spectral lines. The loop is clearly visible in the \sivii~275.36~\AA\ image where the square marks are not shown. The overplotted black-line squares are the $3 pixels \times 3 pixels$ regions where the emission is summed for further analysis.  The blue-line squares mark the regions from which the background emission is obtained. }
\label{t1}
\end{figure*}

Figure~\ref{t1} shows the EIS intensity images of the loop in the  \mgvi\ 268.99~\AA, \mgvii\ 278.40~\AA, \mgvii\ 280.74~\AA, \sivii\ 275.36~\AA\ and \feviii\ 186.60~\AA\ lines that were used for the spectroscopic plasma diagnostics. The loop structure is outlined by the black squares in these intensity maps apart from the EIS~\sivii\ image. The spectral line pair \mgvii\ 278.40~\AA\ and \mgvii\ 280.74~\AA\ was the only line pair suitable for the electron density diagnostics of this loop. To obtain the intensity of \mgvii\ 278.40~\AA, we removed the blend by the \sivii\ 278.44~\AA\ line. To estimate the contribution of \sivii\ 278.44~\AA\ we used the theoretical ratio of \sivii\ 278.44~\AA\ and 275.36~\AA\ \citep{2007PASJ...59S.857Y}. The blue squares plotted along the loop (Figure~\ref{t1}) are the regions where the background emission was obtained and later subtracted from the emission in the loop.

\begin{figure*}[!ht]
\centering
\includegraphics[trim=0cm 0cm 0cm 0cm,clip,width=1.\columnwidth]{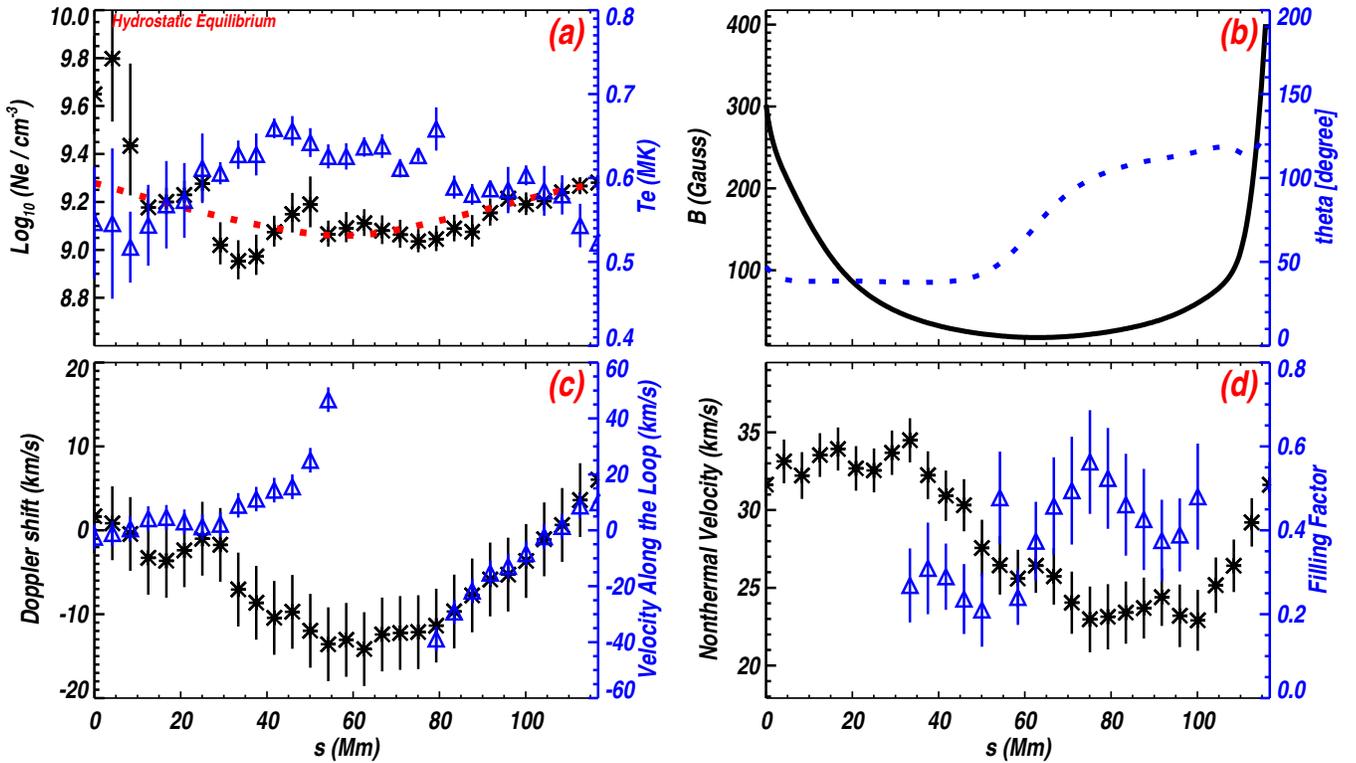}
\caption{Physical parameter distribution along the length of Loop7. In panel (a) black asterisks denote the electron density distribution and the blue triangles the  temperature distribution.  In panel (b) black line is the magnetic flux intensity and the blue dotted line is the $\theta$ distribution. Panel (c) black asterisks show the Doppler velocity distribution and blue triangles  the velocities along the loop. Panel (d) black asterisks show the non-thermal velocity distribution and blue triangles the filling factor distribution.}
\label{tn1}
\end{figure*}

\begin{figure*}[!ht]
\centering
\includegraphics[trim=0cm 0cm 0cm 0cm,clip,width=1.\columnwidth]{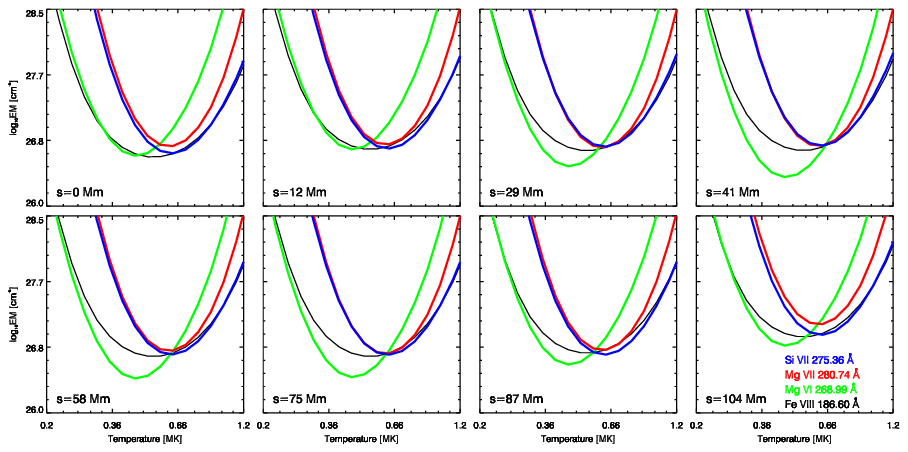}
\caption{Emission measure curves for Loop7.  $s=0$ Mm stands for the loop east footpoint and $s=104$ Mm is the loop west footpoint.}
\label{em1}
\end{figure*}

\par
Figure~\ref{tn1} shows the distribution of the obtained physical  plasma  parameters along the loop length.
Full details on all loop (both cool and warm) parameters are given in Table~\ref{parameters}.
The length of the loop is 117~Mm, the height (the distance from the loop top to the solar surface) is 18~Mm,
   and the averaged radius $\sim$2.3~Mm (see Section~3 on how these parameters were estimated).
The loop electron density reads 10$^{9.65}$~cm$^{-3}$ at the east footpoint, and decreases
    to $10^{9.05}$~cm$^{-3}$ at the loop top (Figure~\ref{tn1}(a)). One can also see an increase in the density from the loop top towards the west footpoint, although not as significant.
Please note that the loop top is the middle along the loop length as determined from the magnetic field extrapolation.
We found a good match between the observed density distribution and the form given by Eq.~(\ref{equathy}) except for
   the part near the east footpoint.

Four spectral lines, i.e., \mgvi\ 268.99~\AA, \mgvii\ 280.74~\AA, \sivii\ 275.36~\AA\ and \feviii\ 186.60~\AA, were used to obtain the electron temperatures of the loop using the EM-loci method. Figure~\ref{em1} shows the EM curves of randomly selected points along the loop. We found that the curves almost cross in a single point which permits us to assign the value in this point as the loop temperature value. We can conclude that the loop is close to being isothermal along the LOS for each of the analysed points along the loop. The loop electron temperatures for Loop7 range from 0.5~MK to 0.65~MK and show an increase from the loop footpoints to the loop top.

\par
The magnetic field along the loop is 302~G in the east footpoint, 18~G at the loop top and 397~G in the west footpoint (Figure~\ref{tn1}(b)). As shown by \citet{2010ApJ...723L.185W}, the magnetic field strength is different in two loop footpoints.  The $\theta$ distribution shows a smooth change from 40 degrees at the east footpoint to 130 degrees at the west footpoint.  The Doppler shifts near the two footpoints of the loop are red-shifted by 2 -- 3~\kms\ and blue-shifted by up to $\sim$15~\kms\ at the loop top (Figure~\ref{tn1}(c)).  Please note that EIS Doppler  velocities for data taken in a rastering mode have errors of $\sim$4.4~\kms\ \citep{2010SoPh..266..209K}. The calculated velocities along the loop length suggest upflow in both footpoints. To explain this, we have made further detailed analysis of all AIA images  taken (at 12~s cadence) during the rastering of this loop. We found that at least two strands comprise the EIS loop. During the time the EIS slit was rastering the western part of the loop (EIS always scans in West -- East direction), one strand dominated the line-of-sight up to a length of  84~Mm (as measure starting in the west footpoint). While the slit was moving further East another strand prevailed along the line-of-sight. Therefore, we believe the interpretation of the proper flows derived from the Doppler-shift measurements of this EIS loop should be taken with caution. We have estimated a strand width from the AIA images and found values of 3.2\arcsec\ -- 3.9\arcsec\ (2.3~Mm  -- 2.8~Mm) which are comparable to the recent finding by \citet{2013A&A...556A.104P} from Hi-C (0.3 -- 0.4\arcsec\ spatial resolution)  data (1.8\arcsec\ -- 2.4\arcsec\ or 1.3~Mm -- 1.7~Mm). \citet{2013ApJ...772L..19B} found that Hi-C loops have a loop-width distribution that peaks at about 270~km. A loop width of 4.6~Mm measured here means that indeed several strands compose the EIS loop. The non-thermal velocities obtained from the \sivii\ 275.36~\AA\ line along the loop range from $\sim$ 23~\kms\ to $\sim$ 35~\kms\ (Figure~\ref{tn1} (d)). The filling factor remains relatively constant for the loop positions where the radius could be estimated best and its averaged value is $\sim$0.38. This filling factor value suggests that the EIS loop is made of unresolved strands (as already indicated by the AIA imaging data). The AIA data reveal as mentioned above that at least 2 strands fill the EIS loop volume.
\begin{center}
\begin{figure*}[!h]
\begin{tabular}{cc}
\hspace{2cm}
\includegraphics[width=.64\textwidth]{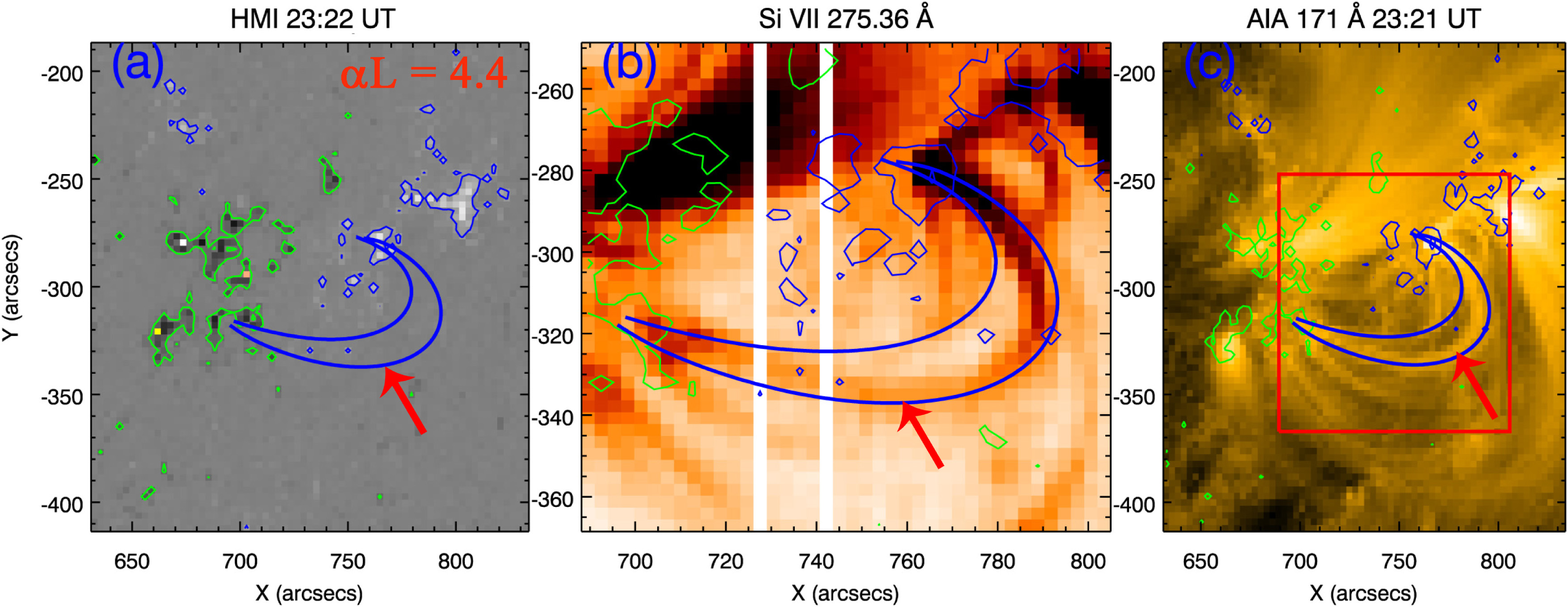} &
\includegraphics[width=.30\textwidth]{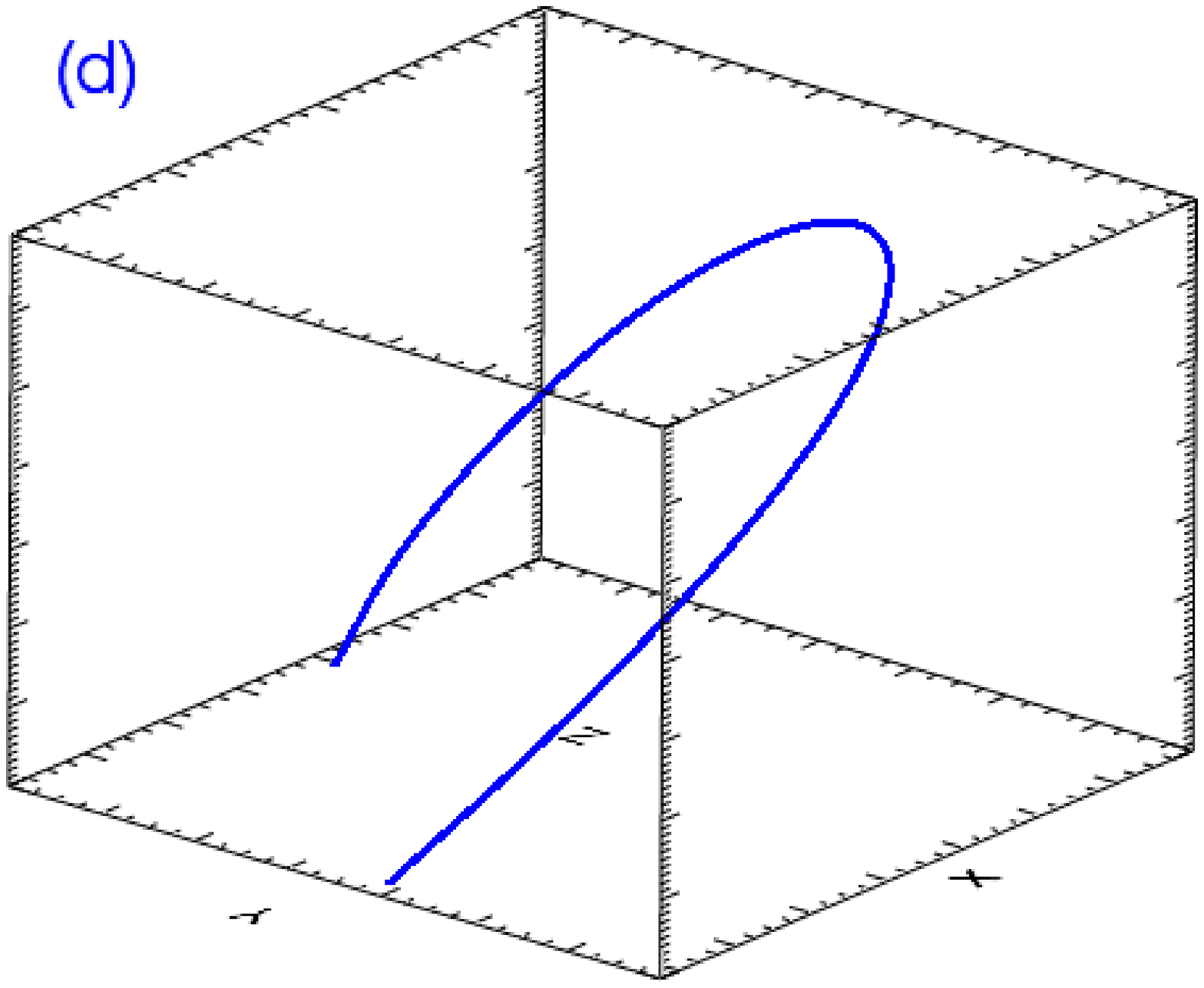}
\end{tabular}
\caption{The same as in Figure~\ref{lfff1} for the second cool loop, Loop12, observed on 2011 February 6.}\label{lfff2}
\end{figure*}
\end{center}

\begin{figure*}[!ht]
\centering
\includegraphics[trim=0cm 0cm 0cm 0cm,clip,width=1.\columnwidth]{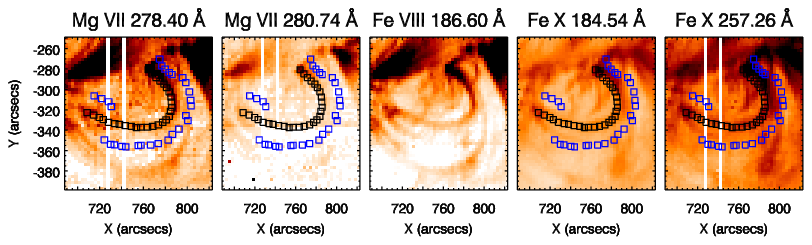}
\caption{The same as Figure~\ref{t1} for Loop12.}
\label{t2}
\end{figure*}
\begin{figure*}[!ht]
\centering
\includegraphics[trim=0cm 0cm 0cm 0cm,clip,width=1.0\columnwidth]{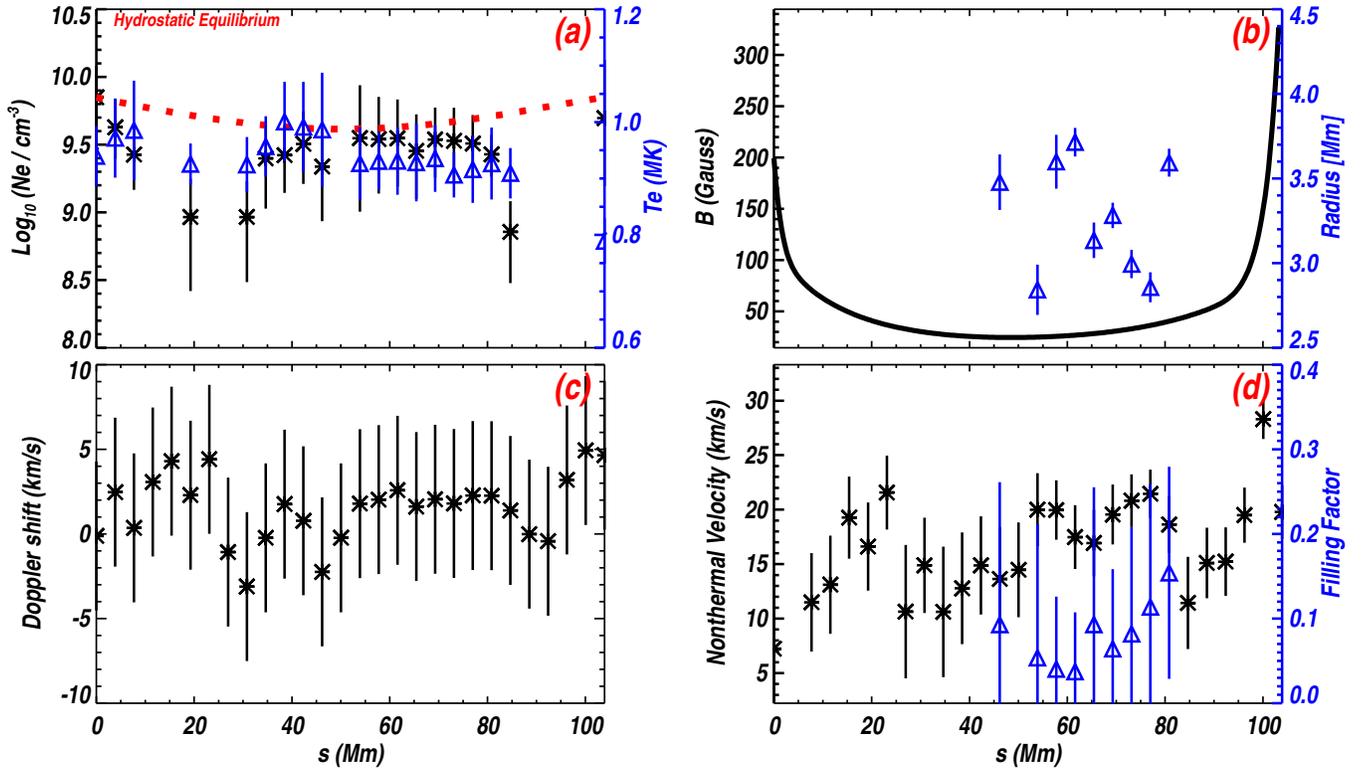}
\caption{The same as Figure~\ref{tn1} for Loop12. Note that the radius distribution along the loop is shown in panel (b) with blue triangles.}
\label{tn2}
\end{figure*}

\subsubsection{Loop12}
Figure~\ref{lfff2}(a) shows that  two extrapolated magnetic-field lines  are in agreement with one fully detectable loop (Loop12) in the EIS raster data (Figure~\ref{lfff2}(b)) and  a second one that is only partially visible in some of the spectral lines. Both loops are also seen in the AIA images (Figure~\ref{lfff2}(c)). Loop12 was fully observed in the \mgvii\ 278.40~\AA, \mgvii\ 280.74~\AA, \feviii\ 186.60~\AA, \fex\ 184.54~\AA, and \fex\ 257.26~\AA\ lines displayed in Figure~\ref{t2}, which indicates that it has higher temperatures in comparison with Loop7. The length, height and radius of the loop are 104~Mm, 30~Mm, and 3.2~Mm, correspondingly.  The 3D view of the  extrapolated magnetic-field line (Figure~\ref{lfff2}(d)) reveals a highly tilted towards the solar surface loop.

Figure~\ref{tn2} shows the plasma parameter distributions along the loop. The electron densities have a bit larger values (with respect to Loop7) of $10^{9.85}$~cm$^{-3}$ -- $10^{9.45}$~cm$^{-3}$ from the east footpoint towards the loop top. An inspection of Fig.~\ref{tn2} (a) indicates that the spatial dependence of the measured electron density
   is way more complex than the simple exponential form as given by Eq.~(\ref{equathy}).

We found that the Loop12 temperature range is  0.75~MK -- 0.99~MK (see Table~\ref{parameters}). The Loop12 magnetic field distribution shows  values of 199, 24, and 324~G in the east footpoint, loop apex and west footpoint, respectively. The Doppler shifts along Loop12 are negligible given the orientation of the loop with respect to the observer. They have small values that range between $-$4~\kms\ and  5~\kms\  and can be assumed zero within the one sigma errors. The non-thermal velocities obtained from \sivii\ 275.36~\AA\ along Loop12 are $\sim$ 15~\kms\ in average. The filling factor for Loop12 has a smaller average value of $\sim$0.08 which is probably due to the higher electron densities.

\subsubsection{Comparison with previous studies of cool loops}

There are only limited  reports on cool loop plasma parameters \citep[e.g.,][]{2003A&A...406.1089D} and only one on a full cool loop by \citet{2015ApJ...800..140G}.  In the analysis by  \citet{2015ApJ...800..140G}, the electron temperatures showed a flat distribution from the east footpoint to the loop top and decreased to the west footpoint from 0.76~MK to 0.58~MK. They found an electron density range of $10^{9.5}$~cm$^{-3}$ -- $10^{8.5}$~cm$^{-3}$ from the two footpoint to the loop top. The density distribution showed faster decrease compared to the cool loops in our analysis, while displayed over-dense plasma in one of the loop footpoints and under-dense in the other under a hydrostatic equilibrium assumption.

\subsection{Warm Loops}

In this study, 39 whole warm loops were identified and analyzed. We only display the figures of Loop22 and Loop29 in the main text.  In the following, we present the results of these loops.

\begin{center}
\begin{figure*}
\begin{tabular}{cc}
\includegraphics[width=.64\textwidth]{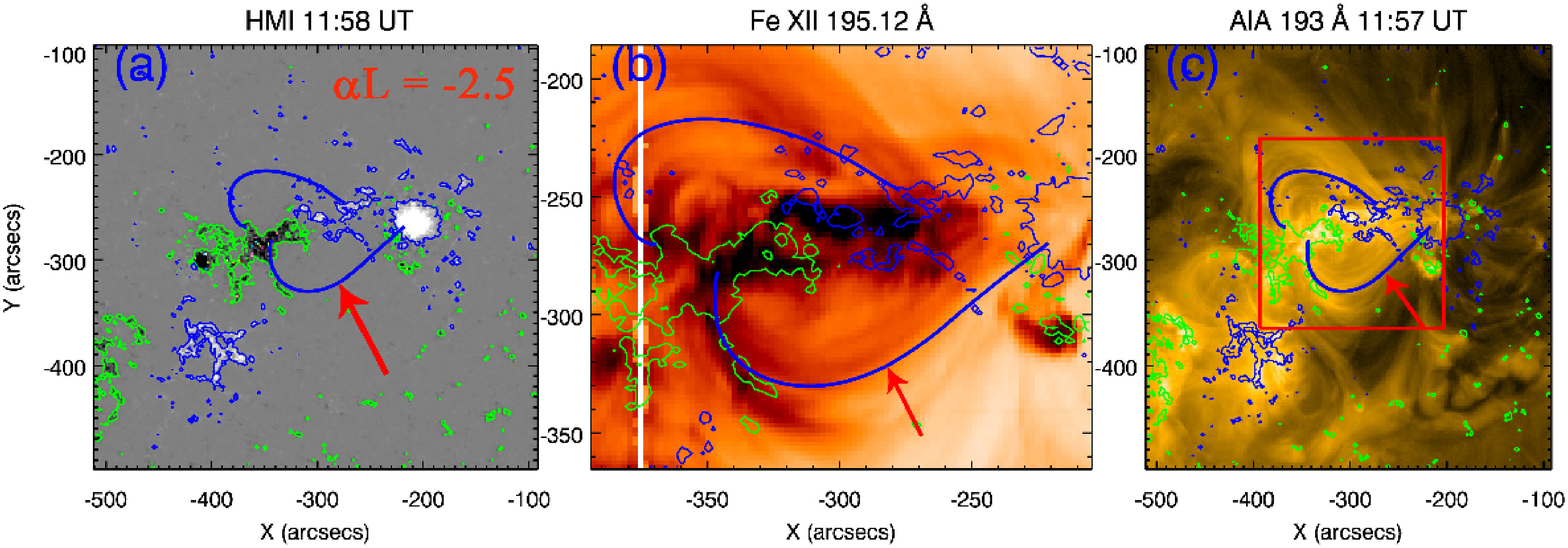} &
\includegraphics[width=.30\textwidth]{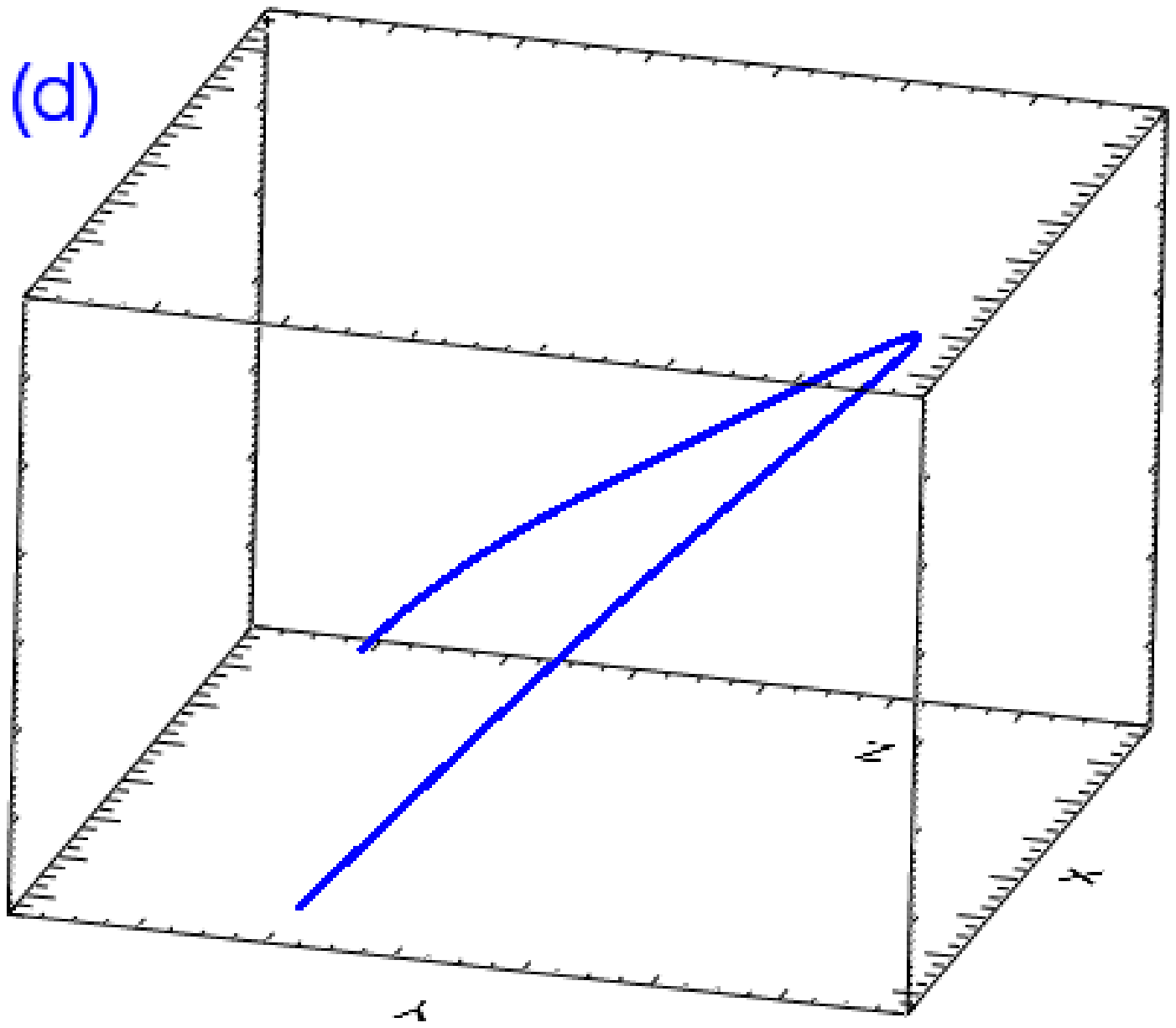}
\end{tabular}
\caption{The same as in Figure~\ref{lfff1} for a warm loop (Loop22) observed on 2012 March 28. The EIS intensity image is taken in the \fexii~195.12~\AA\ line.}\label{lfff5}
\end{figure*}
\end{center}

\begin{figure*}[!ht]
\centering
\includegraphics[trim=0cm 0cm 0cm 0cm,clip,width=1.0\columnwidth]{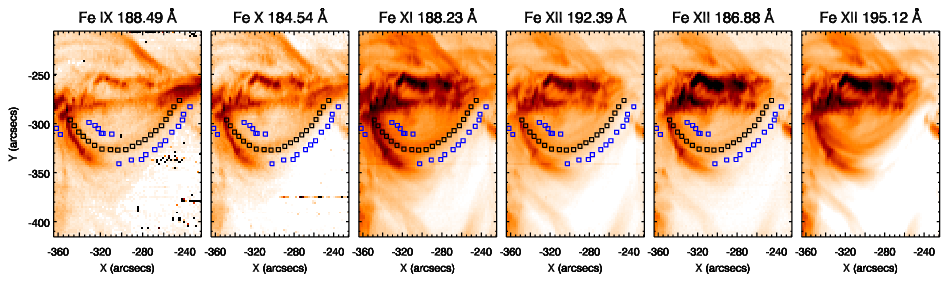}
\caption{The same as Figure~\ref{t1} for Loop22.}
\label{t5}
\end{figure*}

\begin{figure*}[!ht]
\centering
\includegraphics[trim=0cm 0cm 0cm 0cm,clip,width=1.0\columnwidth]{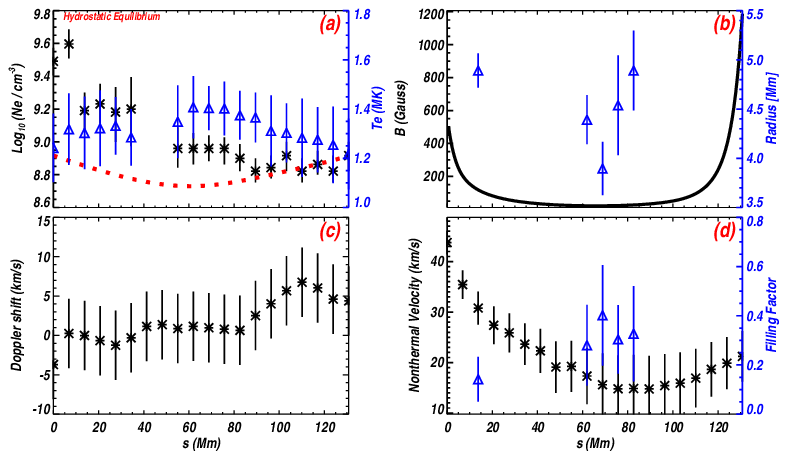}
\caption{The same as Figure~\ref{tn2} for warm Loop22.}
\label{tn5}
\end{figure*}

\begin{figure*}[!ht]
\centering
\includegraphics[trim=0cm 0cm 0cm 0cm,clip,width=1.\columnwidth]{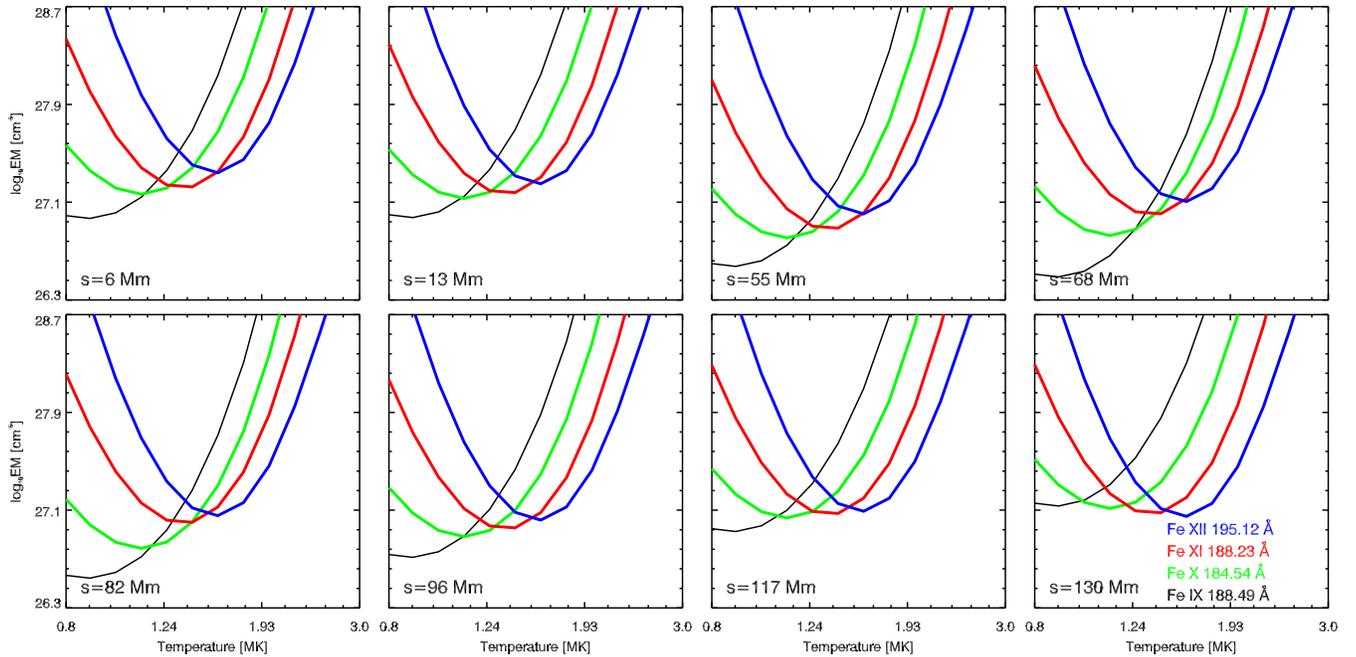}
\caption{Emission measure curves for Loop22.}
\label{em5}
\end{figure*}
\subsubsection{Loop22}

Figure~\ref{lfff5}(a) displays two extrapolated magnetic field lines ($\alpha$L = $-$2.5) overplotted on an HMI magnetogram
  of NOAA 11445 observed on 2012 March 28.
These two lines correspond to only one fully visible loop in the EIS intensity images (Figure~\ref{lfff5}(b)),
   while the second has only one (west) visible leg. Both loops are  distinguishable in AIA~193~\AA\ Figure~\ref{lfff5}(c).
The 3D presentation of Loop22 is given in Figure~\ref{lfff5}(d).
Images in six EIS spectral lines, i.e. \feix\ 188.49~\AA, \fex\ 184.54~\AA, \fexi\ 188.23~\AA, \fexii\ 186.88~\AA, \fexii\ 192.39~\AA, and \fexii\ 195.12~\AA\ of Loop22 are displayed in Figure~\ref{t5}.
Figure~\ref{tn5} presents the physical parameter distributions along the loop.
We found that the electron density decreases from $\sim$10$^{9.5}$~cm$^{-3}$ in the loop east footpoint to $\sim$10$^{8.95}$~cm$^{-3}$ at the loop top.
{The density profile differs considerably from the one described by Eq.~(\ref{equathy}), given by the red dashed curve in Fig.~\ref{tn5} (a).}
Figure~\ref{em5} shows the EM curves for several points along the loop. The temperature in the two loop footpoints has the same value of 1.25~MK. From the west loop footpoint to the loop top, the temperature slightly increases from 1.25~MK to 1.45~MK. The magnetic-field intensity distribution along the loop shows  492~G in the east footpoint,  20~G at the loop top and 1186~G in the west footpoint. We found that the Doppler shifts are close to zero along the loop because of  its orientation towards the observer. The non-thermal velocity values obtained from \fexii\ 195.12~\AA\ are in the range of 17 -- 43~\kms. The filling factor average value is $\sim$0.29 around the loop top. As a reminder for the reader, the filling factor values were only estimated for loop points where reliable radius values of the EIS loop can be estimated.

\subsubsection{Loop29}

\begin{center}
\begin{figure*}[!h]
\begin{tabular}{cc}
\includegraphics[width=.65\textwidth]{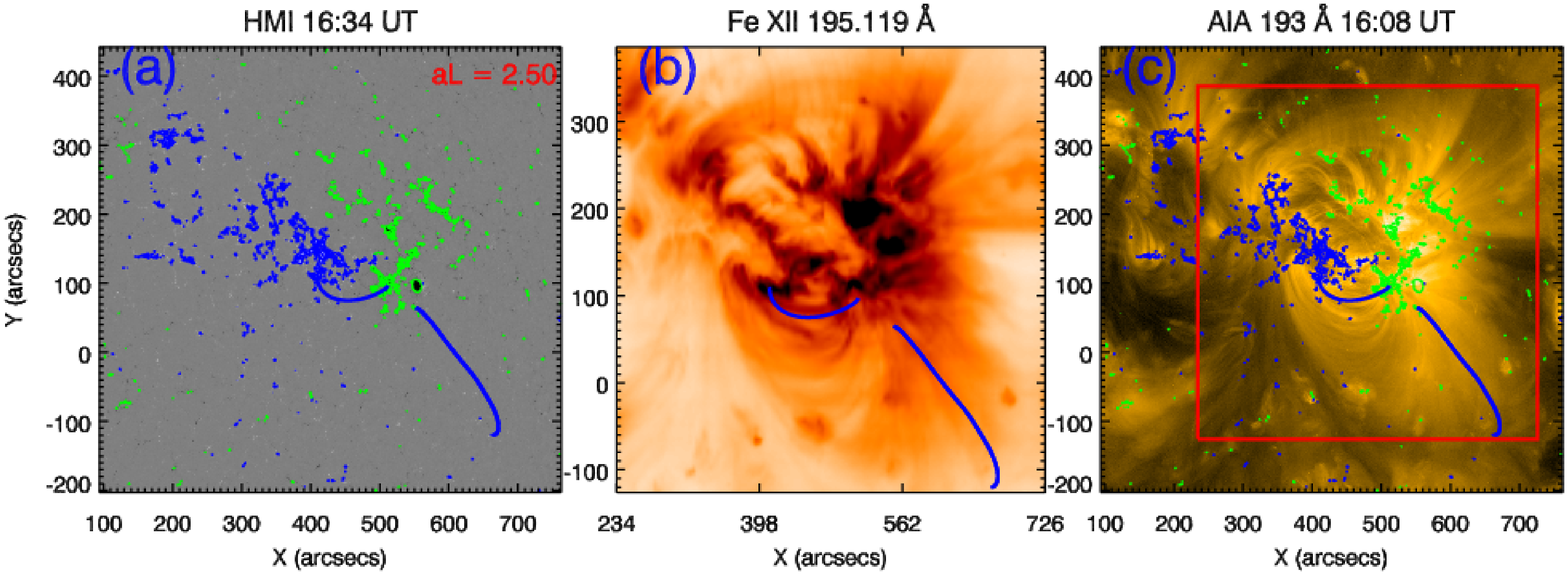} &
\includegraphics[width=.30\textwidth]{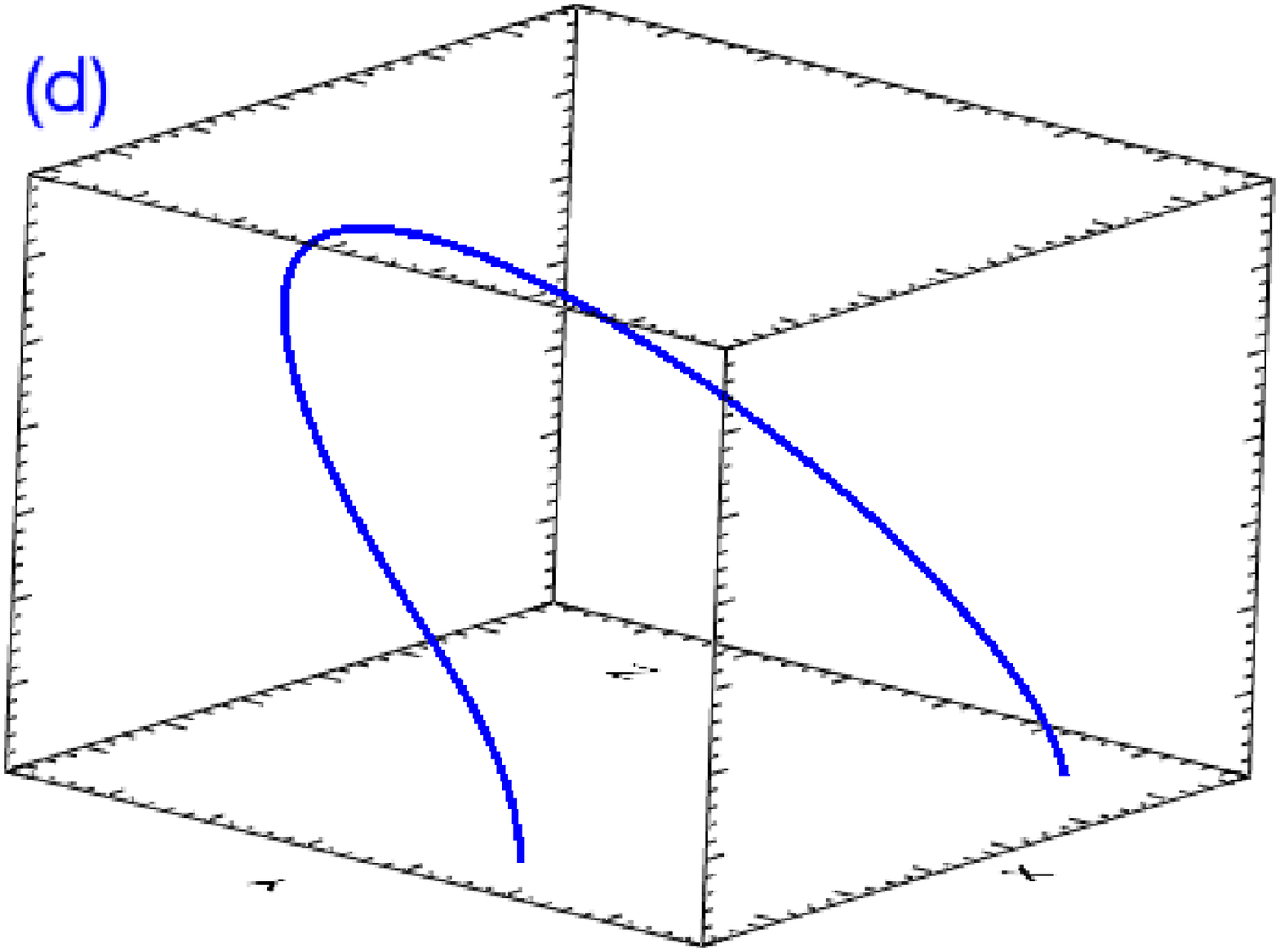}
\end{tabular}
\caption{The same as Figure~\ref{lfff1} for a warm Loop29 observed on 2013 August 8. The EIS intensity image is taken in the \fexii~195.12~\AA\ line.}\label{lfff4}
\end{figure*}
\end{center}

Figure~\ref{lfff4}(a) shows the HMI magnetogram of the observed AR NOAA 11809 registered on 2013 August 8. Two extrapolated magnetic-field lines are identified at $\alpha$L = 2.5 that best agree with two loops. Only one loop, however,  is clearly  visible along its full length in the EIS   \fexii\ 195.12~\AA\ intensity image (Figure~\ref{lfff4}(b)) and was numbered as Loop29. We show  in Figure~\ref{lfff4}(c) the AIA 193~\AA\ loop observations of both loop. The Loop29 3D positioning derived from the extrapolation can be seen in the Figure~\ref{lfff4}(d).

\par

\begin{figure*}[!ht]
\centering
\includegraphics[trim=0cm 0cm 0cm 0cm,clip,width=1.\columnwidth]{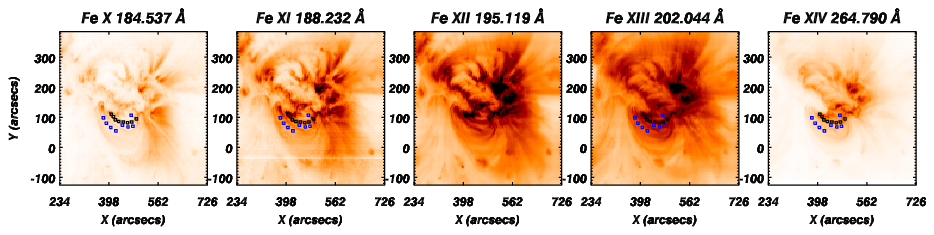}
\caption{The same as Figure~\ref{t1} for warm Loop29.}
\label{t4}
\end{figure*}

\begin{figure*}[!ht]
\centering
\includegraphics[trim=0cm 0cm 0cm 0cm,clip,width=1.\columnwidth]{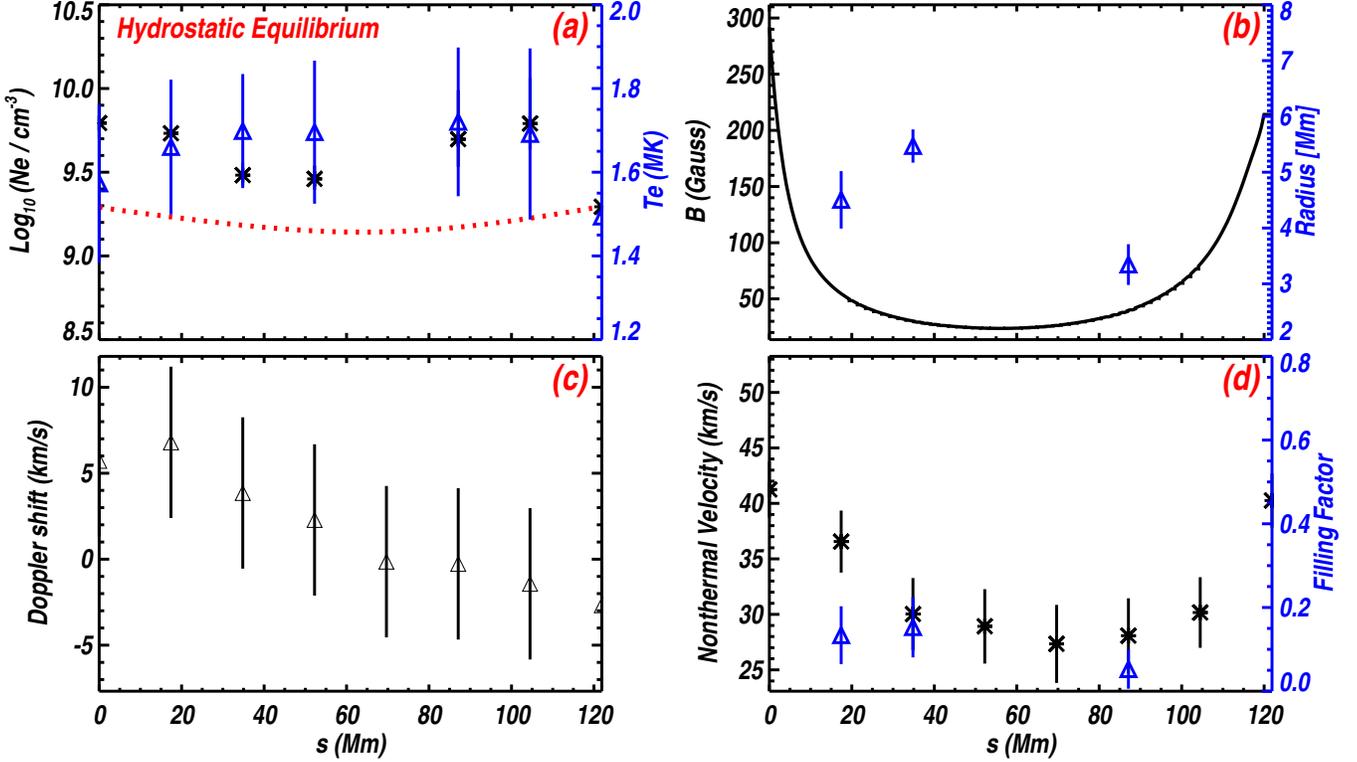}
\caption{The same as Figure~\ref{tn2} for warm Loop29.}
\label{tn4}
\end{figure*}

The loop was registered in the EIS \fex\ 184.54~\AA, \fexi\ 188.23~\AA, \fexii\ 195.12~\AA, \fexiii\ 202.04~\AA, and \fexiv\ 264.79~\AA\ lines (Figure~\ref{t4}). The spectral line pair \fexiii\ 202.04/(203.80+203.83)~\AA\ was used to obtain the electron density, and all spectral lines shown in Figure~\ref{t4} were used to derive the electron temperature. Figure~\ref{tn4} shows the derived physical parameter distributions along Loop29. The loop length is 121~Mm and its height is 34~Mm.  The electron density distribution shows a slightly decrease from the loop east footpoint to up to 60~Mm.  The electron density values range from $10^{9.79}$~cm$^{-3}$ at the loop east footpoint to $10^{9.45}$~cm$^{-3}$ at the loop-top.
{The observed electron density distribution is inconsistent with the exponential form given by Eq.~(\ref{equathy}).}
The temperature distribution along Loop29 that ranges from 1.49~MK  to 1.75~MK. Figure~\ref{tn4}(b), shows the magnetic-field intensity distribution along Loop29. The magnetic field values range from 23~G at the loop top to 291~G in the loop east footpoint, and 214~G in the west. The Doppler shifts along the loop are close to zero because of the orientation of the loop with respect to the observer (see Figure~\ref{lfff4}(d)). The non-thermal velocities obtained from \fexii\ 195.12~\AA\ along the loop range from 27~\kms\ to 41~\kms. The  filling factor values range from 0.05 to 0.14.

\subsubsection{Previous studies on warm loops}

There are several studies  on warm loops (see the Introduction), however only a handful of them  investigate  whole loops. \citet{1998Natur.393..545P} reported that the temperature distribution  of a warm loop observed in SXT data increases from 1.5~MK  in the two footpoints to 2.2~MK  at the loop top. \citet{2004ApJ...608.1133L} found that the electron density distribution increases from $10^{8.9}$~cm$^{-3}$ in the loop footpoints to  -- $10^{9.4}$~cm$^{-3}$ at the loop top. The temperature distribution was constant along the loop and had an  average value of $\sim$1.8~MK. The loop length in their analysis was $\sim$150~Mm obtained from images after a  correction for the loop geometry. \citet{2008ApJ...680.1477A} obtained temperature and density distributions of 7 whole loops from a triple-filter analysis of  stereoscopic images observed with the twin STEREO/EUVI satellite  and gave the statistical results of their physical parameters. For example, loop lengths obtained from a triangulation method were in the range of 60~Mm -- 266~Mm and loop widths $\sim$2.6$\pm$0.1~Mm which after correcting for the instrumental point-spread function correspond
to effective loop widths of 1.1  $\pm$ 0.3~Mm. The temperatures along loops were found ``to
be nearly constant, within the uncertainties of the background subtraction''. The electron densities in loop footpoints were determined as $10^{9.2}$~cm$^{-3}$ -- $10^{9.5}$~cm$^{-3}$ and the distributions show agreement with hydrostatic equilibrium state. Next, we compare the physical parameters determined in earlier studies with our results.
\begin{figure*}[!ht]
\centering
\includegraphics[trim=0cm 0cm 0cm 0cm,clip,width=0.8\columnwidth]{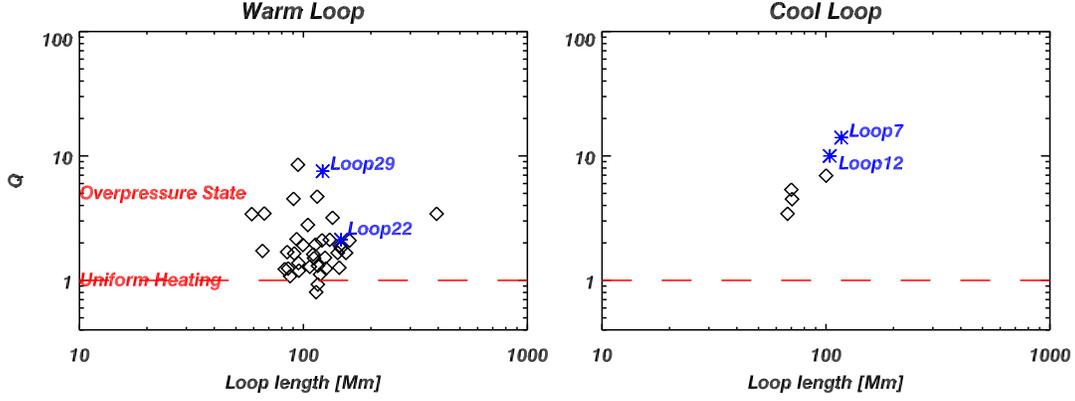}
\caption{Loop pressure ratio Q ($\bar{P}$ to P$_{RTV}$) vs. the loop length L for warm (left panel) and cool (right panel) loops.
The cool and warm loops presented here in detail are marked blue. }
\label{statistics}
\end{figure*}
\subsection{Pressure states of cool and warm loops}

Given the relatively large size of the loop sample that we analyze,
   we may now proceed to examine how our loops behave in a statistical sense by comparing their parameters with
   the well-known RTV law~\citep[][hereafter RTV78]{1978ApJ...220..643R}.
In addition to the loops being static, RTV$78$  assumed that thermal pressure is a constant, and that
   the energy balance is achieved among the electron heat flux, radiative cooling, and a uniform heating
   rate.
The end result is that, in cgs units, the loop pressure $\mathrm{P_{\mathrm{RTV}}}$ can be expressed as
\begin{equation}
   \mathrm{P_{RTV} \approx \frac{2}{L} \bigg(\frac{T_{max}}{1400}\bigg)^3},
\label{equat8}
\end{equation}
 where $\mathrm{T_{\mathrm{max}}}$ is the maximum temperature along a loop, and the loop length $\mathrm{L}$ is derived from our force-free field extrapolation.
Several studies have tested the RTV scaling law and found a discrepancy between the theoretical and the observational results \citep[e.g.,][and the references therein]{1981ApJ...247..692P, 2008ApJ...680.1477A,2016A&A...589A..86B}.

Following \citet{2008ApJ...680.1477A}, we quantify the departure of our loop parameters from the RTV law by evaluating
   the so-called overpressure ratio
\begin{equation}
  \mathrm{ Q = \frac{P_{\mathrm{obs}}}{P_{\mathrm{RTV}}} = \frac{\bar{P}}{(2/L)(T_{\mathrm{max}}/1400)^{3}}}~,
\label{equat9}
\end{equation}
where the loop pressure $\mathrm{P_{\mathrm{obs}}}$ is obtained as  $\mathrm{2 n_{e} k_{\mathrm B} T_{e}}$  and an average value  $\mathrm{P_{\mathrm{obs}}}$  is derived ($\bar{P}$) from segments of the loop
   where the pressure does not vary significantly.
Here, $\mathrm{n_e}$ and $\mathrm{T_e}$ are the measured electron density and temperature (using the EM-loci method), respectively.
Furthermore, $\mathrm{k_{\mathrm B}}$ denotes the Boltzmann constant.
The factor $2$ comes from the assumption that the loop plasma comprises primarily electrons and protons.

Figure~\ref{statistics} shows the overpressure ratio $\mathrm{Q}$ as a function of loop length $\mathrm{L}$ for warm (the left panel) and cool (right) loops.
Note that $6$ out of our $50$ loops were not included because no suitable density-sensitive line pairs were available for us to derive
    the electron densities.
One sees that the majority of the warm loops show an overpressure ratio in the range $1.2 - 8.5$.
As a matter of fact, $33$ out of our $38$ warm loops fall in this range.
As for the rest, $5$ loops correspond to a $\mathrm{Q}$ in the range $0.8-1.2$.
When it comes to the 6 cool loops we examine, the derived values of $\mathrm{Q}$ are exclusively larger than $2$.
In any case, Figure~\ref{statistics} suggests that neither warm nor cool loops that we examine agree with
   the scaling law predicted by RTV78.
We note that a similar discrepancy was also found for warm loops such as measured with STEREO by \citet{ 2008ApJ...680.1477A}.

\begin{center}
\begin{longtable}{c c c c c c c c c c}
\caption{Physical parameters of the loops.}\label{parameters}\\
\hline

Loop&Temperature&Density&Length&Height&Radius&Filling factor&$V_{nt}$\tablenotemark{a} & B&$\alpha$L \\
No.(Date)&(ef$/$top$/$wf)&(ef$/$top$/$wf)&&&&&&(ef$/$top$/$wf)&\\
&[MK]&[log$_{10}cm^{-3}$]&[Mm]&[Mm]&[Mm
]&&[{\rm km\;s$^{-1}$}]&[Gauss]&\\ \hline
Lp1& $1.26/1.63/1.72$ & $8.93/8.77/9.39$ &147&16&$ 2.5 \pm 0.17 $ &$ 0.30 \pm 0.16 $ & $28 - 38$&$137/11/91$&-1.5\\
Lp2 [C]& $0.85/0.77/0.93$ &*&62 &11&$ 1.1 \pm 0.05 $&*& $29 - 44$ &$242/23/558$&-4.0 \\
Lp3 [C]& $0.84/0.88/0.95$ & $9.17/9.10/9.01$ &70&20&$ 3.5 \pm 0.32 $ &$ 0.21 \pm 0.20 $ & $12 - 35$&$506/23/219$&4.0\\
Lp4 [C]& $0.92/0.89/0.84$ & $9.88/9.11/9.43$ &100 &19&$ 2.6 \pm 0.16 $&$ 0.19 \pm 0.07 $& $13 - 32$ &$70/7/95$&4.0 \\
Lp5 [C]& $0.83/0.96/0.94$ & $9.04/9.14/10.40$ &67&18&$3.8 \pm 0.18$&$ 0.45 \pm 0.14 $ & $10 - 30$  &$746/23/135$&4.0\\
Lp6 [C]& $0.91/0.98/0.85$ & $9.40/9.01/9.26$ &70&21&$ 2.7 \pm 0.11$ &$ 0.35 \pm 0.19 $ &  $10 - 40$  &$270/19/258$&3.5\\
Lp7 [C]& $0.55/0.65/0.52$ & $9.65/9.05/9.25$ &117 &18&$ 2.3 \pm 0.08 $&$ 0.38 \pm 0.03 $& $23 - 35$ &$302/18/397$&0 \\
Lp8& $1.85/1.95/1.71$ & $9.04/8.81/9.11$ &118&18&$ 3.9 \pm 0.18$ &$ 0.84\pm 0.17 $& $3 - 35$  &$296/30/184$&2.5\\
Lp9& $1.79/1.89/1.53$ & $9.60/9.20/9.17$ &154&42&$ 3.7 \pm 0.26 $ &$ 0.45 \pm 0.16 $& $13 - 35$  &$234/33/1541$&-2.0\\
Lp10 [C]& $0.74/0.89/0.79$ & * &196&26&$ 1.4 \pm 0.11 $ &*& $20 - 47$&$300/28/1293$&-4.0\\
Lp11& $1.46/1.55/1.64$ & *&125 &17&$ 1.7 \pm 0.11 $&*& $10 - 33$ &$510/12/329$&-1.0 \\
Lp12 [C]& $0.95/0.99/0.75$ & $9.85/9.45/9.75$ &104&30&$ 3.2 \pm 0.11 $ &$ 0.08 \pm 0.04 $ & $7 - 28$&$199/24/324$&4.4\\
Lp13 [C]& $1.07/1.05/1.02$ &*&148&40&$ 2.4 \pm 0.12$ &*&  $22 - 32$  &$120/15/317$&0\\
Lp14& $1.55/1.50/1.55$ & $9.60/8.95/9.60$ &91&15&$ 3.1 \pm 0.16 $ &$ 0.25 \pm 0.08 $& $18 - 38$  &$465/53/290$&-1.0\\
Lp15& $1.55/1.75/1.55$ & $9.65/8.75/9.65$ &116&17&$ 3.2 \pm 0.08$ &$ 0.31\pm 0.08 $& $18 - 39$  &$193/33/346$&-1.0\\
Lp16& $1.71/1.54/*$ & $9.17/9.19/10.12$ &113&12&$ 3.5 \pm 0.25$ &$ 0.29\pm 0.14 $& $28 - 36$  &$198/23/278$&-0.5\\
Lp17& $1.46/1.66/1.50$ & $*/8.97/9.62$ &106&15&$ 3.1 \pm 0.18 $ &$ 0.42 \pm 0.14 $& $30 - 35$  &$83/25/297$&-0.5\\
Lp18& $1.77/1.73/1.74$ & $8.96/8.60/9.05$ &142 &30&$ 3.2 \pm 0.12 $ &$0.26 \pm 0.28$ & $13 - 40$ &$386/10/194$&0.5 \\
Lp19& $1.61/1.72/1.75$ & $8.86/8.67/9.17$ &147&55&$4.5 \pm 0.60$&$ 0.20 \pm 0.18 $ & $20 - 33$  &$857/12/747$&0\\
Lp20& $1.14/1.41/2.12$ & $9.49/8.75/*$ &116 &18&$2.8 \pm 0.20$ &$ 0.89\pm 0.34 $ &$20 - 48$   &$60/29/40$&3.5\\
Lp21& $1.58/1.75/1.45$ & $9.11/9.20/8.84$ &111&32&$ 3.5 \pm 0.11$ &$ 0.47 \pm 0.21 $ & $32 - 37$&$194/23/274$&3.0\\
Lp22& $1.25/1.45/1.25$ & $9.50/8.95/8.95$ &131&34&$ 4.5 \pm 0.32$ &$ 0.29 \pm 0.07 $ & $17 - 43$&$492/20/1186$&-2.5\\
Lp23& $1.51/1.70/1.65$ & $9.25/8.95/9.25$ &95&19&$ 3.2 \pm 0.07$ &$ 0.58 \pm 0.22 $ &  $22 - 31$  &$502/10/141$&-4.0\\
Lp24& $1.34/1.80/1.58$ & $9.50/8.80/9.05$ &114 &20&$2.9 \pm 0.25$ &$ 0.29\pm 0.10 $ &$17 - 34$   &$50/12/151$&-2.0\\
Lp25& $1.55/1.70/1.75$ & $*/8.90/9.50$ &126&21&$3.8 \pm 0.23$&$ 0.35 \pm 0.18 $ & $17 - 30$  &$649/23/132$&3.5\\
Lp26 [C]& $0.77/0.90/0.94$ & * &138&36&$ 3.4 \pm 0.35 $ &*& $37 - 47$  &$576/28/229$&-1.0\\
Lp27& $*/1.62/1.36$ & $9.87/8.72/8.93$ &393&93&$ 3.9 \pm 0.36 $ &$ 0.43 \pm 0.28 $ & $19 - 37$&$172/4/239$&1.5\\
Lp28& $1.59/1.50/*$ & $9.24/9.25/*$ &115 &11&$ 3.2 \pm 0.23 $&$ 0.21 \pm 0.13 $& $25 - 31$ &$543/21/308$&-1.5 \\
Lp29& $1.57/1.75/1.49$ & $9.79/9.45/9.29$ &121&34&$ 4.5 \pm 0.40$ &$ 0.14 \pm 0.07 $ &  $27 - 41$  &$291/23/214$&2.5\\
Lp30& $1.51/1.73/1.52$ & $9.36/8.80/9.11$ &110 &20&$3.1 \pm 0.13$ &$ 0.54\pm 0.22 $ &$19 - 33$   &$340/18/126$&1.5\\
Lp31& $1.56/1.58/1.60$ & $8.97/8.67/9.10$ &160&20&$ 3.0 \pm 0.16$ &$ 0.58 \pm 0.17 $ & $30 - 44$&$129/12/344$&-1.5\\
Lp32& $1.58/1.68/1.65$ & $9.56/9.67/9.95$ &67&10&$ 2.8 \pm 0.09$ &$ 0.21\pm 0.07 $& $25 - 32$  &$505/94/480$&4.0\\
Lp33& $1.80/1.71/1.72$ & $9.45/8.98/9.77$ &121&36&$ 3.8 \pm 0.31 $ &$ 0.28 \pm 0.09 $& $32 - 36$  &$269/21/613$&0.5\\
Lp34& $1.54/1.58/1.80$ & $9.88/9.15/9.75$ &94 &30&$ 2.6 \pm 0.07 $&$ 0.19 \pm 0.19 $& $28 - 37$ &$309/26/228$&-4.0 \\
Lp35& $1.51/1.61/1.52$ & $9.36/9.23/9.35$ &65&11&$5.8 \pm 0.83$&$ 0.32 \pm 0.10 $ & $21 - 37$  &$304/27/166$&-1.5\\
Lp36& $1.46/1.59/1.51$ & $9.29/9.07/9.13$ &99&12&$ 5.8 \pm 0.11$ &$ 0.60 \pm 0.40 $ &  $24 - 36$  &$360/19/217$&-1.5\\
Lp37& $1.60/1.69/1.56$ & $10.17/9.48/9.82$ &90 &25&$2.7 \pm 0.17$ &$ 0.24\pm 0.26 $ &$24 - 36$   &$200/31/315$&-1.5\\
Lp38& $1.67/1.61/1.57$ & $9.47/9.25/9.44$ &84&15&$ 3.4 \pm 0.22$ &$ 0.30 \pm 0.11 $ & $28 - 40$&$223/48/171$&-4.0\\
Lp39& $1.53/1.64/1.55$ & $9.61/9.25/9.71$ &104&21&$ 2.0 \pm 0.13$ &$ 0.71\pm 0.20 $& $31 - 43$  &$180/30/782$&4.0\\
Lp40& $1.50/1.64/1.46$ & $9.65/9.49/10.11$ &59&14&$ 2.2 \pm 0.15 $ &$ 0.21 \pm 0.12 $& $27 - 48$  &$365/212/954$&-0.5\\
Lp41& $1.64/1.61/1.58$ & $9.45/9.08/9.31$ &93&24&$ 3.5 \pm 0.14 $ &$ 0.59 \pm 0.17 $ & $20 - 38$&$286/34/368$&3.0\\
Lp42& $1.61/1.73/1.58$ & $9.55/8.63/9.42$ &87 &21&$ 4.2 \pm 0.76 $&$ 0.53 \pm 0.14 $& $21 - 37$ &$181/35/383$&3.0 \\
Lp43& $1.58/1.63/1.57$ & $9.41/8.84/9.49$ &83&19&$4.6 \pm 0.31$&$ 0.64 \pm 0.13 $ & $21 - 42$  &$349/40/142$&0\\
Lp44& $1.45/1.89/1.55$ & $9.50/8.75/9.31$ &144&15&$ 3.5 \pm 0.42$ &$ 0.48 \pm 0.21 $ &  $20 - 31$  &$84/14/101$&2.5\\
Lp45& $1.40/1.64/1.55$ & $9.22/8.60/9.19$ &124 &18&$4.7 \pm 0.27$ &$ 0.62\pm 0.33 $ &$16 - 35$   &$202/9/25$&-3.5\\
Lp46& $1.45/1.76/1.40$ & $*/9.01/9.38$ &95&12&$ 4.0 \pm 0.32$ &$ 0.71 \pm 0.34 $ & $20 - 26$&$173/7/63$&-4.0\\
Lp47& $1.39/1.82/1.58$ & $9.36/8.65/9.30$ &85&10&$ 3.3 \pm 0.07$ &$ 0.54\pm 0.28 $& $21 - 32$  &$207/15/68$&-4.0\\
Lp48 [C]& $0.88/0.96/0.66$ & * &123&22&$ 2.1 \pm 0.40 $ &*& $32 - 42$  &$155/19/260$&-2.5\\
Lp49& $1.52/1.85/1.63$ & $9.35/9.04/9.32$ &134&22&$ 4.2 \pm 0.06 $ &$ 0.24 \pm 0.13 $ & $16 - 36$&$604/17/48$&0.5\\
Lp50& $1.55/1.80/1.60$ & $9.65/8.77/9.15$ &145 &15&$ 3.7 \pm 0.47 $&$ 0.68 \pm 0.31 $& $28 - 38$ &$864/17/132$&0.5 \\
\hline
\hline
\end{longtable}
\begin{tablenotes}
        \footnotesize
        \item[1]Note. The `ef' stands for the east footpoint, `top' for the loop top and `wf' for the west footpoint. The referenced ranges for the non-thermal velocities are the minimum and maximum values along the loops. a: Non-thermal velocities.  The radius and the filling factor are the average of the obtained values. [C] denotes the cool loops
        \end{tablenotes}
\end{center}

\subsection{Comparison between previous and our analyses}

\begin{center}
\begin{longtable}{c c c}
\caption{Temperature and density ratios. The electron density ratios are footpoint/top while the electron temperature ratios  are top/footpoint.}\label{ne_te_fatio}\\
\hline

\hline
LoopNo.&$\mathrm{R_{Ne}}$&$\mathrm{R_{Te}}$ \\
\hline
Lp1&2.81 (\fexii)&1.09\\
Lp2&*&0.86\\
Lp3&0.99 (\mgvii)&0.98\\
Lp4&3.99 (\mgvii)&1.01\\
Lp5&9.50 (\mgvii)&1.08\\
Lp6&2.11 (\fexii)&1.11\\
Lp7&2.78 (\mgvii)&1.21\\
Lp8&1.84 (\six)&1.10\\
Lp9&1.72 (\six)&1.14\\
Lp10&*&1.16\\
Lp11&*&1.01\\
Lp12&2.25 (\mgvii)&1.16\\
Lp13&*&1.00\\
Lp14&4.47 (\fexii)&0.97\\
Lp15&7.94 (\fexii)&1.13\\
Lp16&4.73 (\fexii)&0.90\\
Lp17&4.47 (\fexii)&1.12\\
Lp18&2.55 (\six)&0.99\\
Lp19&2.36 (\six)&1.02\\
Lp20&5.50 (\fexii)&0.87\\
Lp21&0.62 (\fexii)&1.16\\
Lp22&2.27 (\fexii)&1.16\\
Lp23&2.00 (\fexii)&1.08\\
Lp24&3.40 (\fexii)&1.23 \\
Lp25&3.98 (\fexii)&1.03\\
Lp26&*&1.05\\
Lp27&7.87 (\fexiii)&1.19\\
Lp28&0.98 (\fexii)&0.95\\
Lp29&1.44 (\fexiii)&1.10\\
Lp30&2.84 (\fexii)&1.14\\
Lp31&2.34 (\fexii)&1.01\\
Lp32&1.34 (\fexii)&1.04\\
Lp33&4.56 (\fexii)&0.97\\
Lp34&4.68 (\fexii)&0.95\\
Lp35&1.33 (\fexii)&1.06\\
Lp36&1.40 (\fexii)&1.07\\
Lp37&3.54 (\fexii)&1.07\\
Lp38&1.60 (\fexii)&0.99\\
Lp39&2.59 (\fexii)&1.06\\
Lp40&2.81 (\fexii)&1.11\\
Lp41&2.02 (\fexii)&1.01\\
Lp42&7.24 (\fexii)&1.08\\
Lp43&4.09 (\fexii)&1.03\\
Lp44&4.63 (\fexii)&1.26\\
Lp45&4.02 (\fexii)&1.11\\
Lp46&2.34 (\fexii)&1.24\\
Lp47&4.80 (\fexii)&1.23\\
Lp48&* &1.25 \\
Lp49&1.97 (\fexii)&1.17\\
Lp50&4.99 (\fexii)&1.14\\
Kano (1996)&*&$3 \sim 5$ \\
Priest (1998)&*&1.34\\
Lenz (1999)&*&1.05\\
Schmelz (2001)&2.0&1.07 (filter-ratio)\\
&&1.90 (DEM)\\
Del Zanna (2003)&6.30&1.32\\
Landi (2004)&0.50&1.15\\
Gupta (2015)&9.8&1.31\\
\hline
\hline
\end{longtable}
\begin{tablenotes}
        \footnotesize
        \item[1]The asterisk sign (*) denotes that no data are available.
        \end{tablenotes}
\end{center}

Table~\ref{ne_te_fatio} lists the electron density and temperature ratio between the loop footpoints and loop tops from our and previous analyses.  Here, the ratios were calculated using the average values of the density and the temperatures in the two footpoints. As a reminder, the loop top is the middle along the loop length as determined from the magnetic field extrapolation. The temperature distribution along the loops plays an important role in testing loop heating models \citep{1998Natur.393..545P}. \citet{1999ApJ...517L.155L} found that the ratio is near to 1.05 and there were no temperature variations along the loops. \citet{1996PASJ...48..535K} obtained in SXT/Yohkoh data  bigger values for hot loops of  3 -- 5 that indicate a strong increase of temperatures  towards the loop tops. In our analysis for both cool and warm loops,  the temperature ratios range from 0.86 to 1.26 suggesting that  the temperature distribution and thus the  heating mechanism is related to the  loop physical conditions, i.e. size, density, magnetic field etc. The temperature distributions are considered as  increasing from the loop footpoints to the loop top when the temperature ratio values are larger than 1.2. From 50 whole loops, 7 show a temperature increase  from the loop footpoints to the loop top which is in agreement with earlier investigations \citep{1996PASJ...48..535K, 1998Natur.393..545P, 2015ApJ...800..140G}. The rest 43 loops, however,  display more flat distributions in comparison with previous studies  \citep{1999ApJ...517L.155L, 2008ApJ...680.1477A}. \citep{2001ApJ...556..896S} reported density ratio values of  2.0 while derived a ratio of 9.8  in the case of a cool loop. The electron density ratios in our analysis range from $\sim$0.62 to $\sim$9.5 with 36 out of 44 whole loops having a ratio of 1--5.

\section{Summary}
\label{sum}
Using  observations from Hinode/EIS, we analyzed 50 whole loops. Eleven loops were classified as cool ($<$1~MK) and 39 as warm (1 -- 2~MK) loops. We measured their plasma parameters such as electron densities and temperatures, filling factors, velocities, and non-thermal velocities. We took special care of the emission background removal in calculating the loop plasma parameters. For the first time we combined spectroscopic analysis with linear force-free magnetic-field extrapolation.  Our main results are listed in Table~\ref{parameters} and can be  summarized as follows:

\par
The 3D structure of the loops has been derived from the extrapolated magnetic field lines together with their lengths and heights. The investigated loops have heights in the range of 10 -- 93 Mm. The lengths of these loops are from 59~Mm to 393~Mm. The cool loops are  similar in size to the warm loops.  We obtained the radii of the loops by fitting the intensity distributions across the loops in position with a sufficient  signal-to-noise ratio. We found that the radii are in the range of 1.1~Mm -- 5.8~Mm. The magnetic field strength shows decrease from the loop footpoints to the loop tops and  it is different at both footpoints as also found by \citet{2010ApJ...723L.185W}. Their values range from 4~G to 212~G at loop tops while the values in the footpoints are from 25~G to 1541~G.
\par
 Most of the loops show nearly close to zero Doppler shifts  because of their orientation towards the observer.  As for the non-thermal velocity, the values are in the range from 3~\kms\ to 48~\kms\ (obtained in the \sivii\ 275.36~\AA\ and \feviii\ 186.60~\AA\ lines for the cool loops and in the \fexii\ 195.12~\AA\ line for the warm loops).  Observed non-thermal velocities were compared with predictions from models for Alfv\'en
wave turbulence in coronal loops by \citet{2014ApJ...786...28A}. Nonlinear force-free field modeling was used to derive the loop length and magnetic field. The non-thermal velocities in their study range from 15 to 30~\kms\ for different coronal lines that are  comparable with our results taking into account that a different approach was used to remove the instrumental width. The study concludes that  the ``Alfv\'en wave turbulence model is a strong
candidate for explaining how the observed loops are heated.''

\par
 We used the spectral line pair Mg~{\sc vii}~278.40/280.74~\AA\ for cool loops and for warm loops the pairs Fe~{\sc xii}~186.88/
 (195.12+195.18)~\AA, Fe~{\sc xiii} 202.04/(203.80+203.83)~\AA, and \ Si~{\sc x}~258.37/261.04~\AA\  to derive the electron densities. The electron density distributions show a downtrend from the loop footpoints to loop tops for most of the loops. We also compared the observed density distributions with the values under the hydrostatic equilibrium state and found that most of the loops are not in a static state.

 \par
The temperatures of the  loop plasmas are obtained using the EM-loci method. We found that the loop temperatures along the LOS are close to isothermal for all loops. There are eleven loops that have  temperature values below 1~MK, while the rest of the loops are warm loops with  temperatures in the range of 1 -- 2~MK. We found that the temperature distributions are slightly uptrend (from loop footpoints to loop tops) or are nearly flat along the loops. The filling factors of the loops are 8\% -- 89\%. These results suggest that the EIS loops are made either of several unresolved strands or a single strand which does not fill the volume spatially resolved by EIS. This has been modelled in the multi-strand loop simulations of \citet{2012ApJ...755L..33B} where assuming 1~km resolution  and isothermal plasma for each thread, a combined profile is produced from the profiles of several threads which  is then convolved with the EIS PSF and re-scaled to the instrument pixel resolution. Their results have demonstrated that ``million degree loops are revealed to be single monolithic structures that are fully spatially resolved by current instruments. The majority of loops, however, must be composed of a number of finer, unresolved threads.'' Our results strongly support their conclusions.

It remains a challenge to understand the heating of coronal loops -- the building blocks of the solar corona \citep{2006SoPh..234...41K}.
Given the size of the loop sample we gathered,
   a detailed analysis of both the plasma and geometrical properties of coronal loops such as presented in the text
   will likely contribute to a more concrete understanding of how coronal loops are heated.
Evidently, modeling efforts will be indispensable in this aspect.
In a follow-up study, we will report on the comparison
   of the measurements with our loop models where the plasma heating derives from
   turbulent Alfv\'en waves~\citep{2006RSPTA.364..533L}.

\acknowledgments
{\it Acknowledgments:}
This research is supported by the
National Natural Science Foundation of China (41474149, 41674172, 41274176, 41404135, 41274178, 41474150).
ZH thanks the Shandong provincial Natural Science Foundation (ZR2014DQ006)
and the China Postdoctoral Science Foundation. TW acknowledges support by DFG-grant WI 3211/4-1.
Hinode is a Japanese mission developed and launched by ISAS/JAXA, collaborating with NAOJ as a domestic partner, NASA and STFC (UK) as international partners. The scientific operations of the Hinode mission are conducted by the Hinode science team organized at ISAS/JAXA. Support for the post-launch operation is provided by JAXA and NAOJ (Japan), STFC (U.K.), NASA, ESA, and NSC (Norway).

\bibliographystyle{aasjournal}
\bibliography{haixia1}

\end{document}